\documentclass[a4paper,fleqn,usenatbib]{mnras}

\usepackage{Times}

\usepackage[T1]{fontenc}
\usepackage{ae,aecompl}

\usepackage{graphicx}	
\usepackage{amsmath}	
\usepackage{amssymb}	

\title[The origin of scatter in the stellar mass - halo mass relation]{The origin of scatter in the stellar mass - halo mass relation of central galaxies in the EAGLE simulation} 

\author[J. Matthee et al.]{Jorryt Matthee$^{1}$\thanks{E-mail: matthee@strw.leidenuniv.nl}, Joop Schaye$^{1}$, Robert A. Crain$^{2}$, Matthieu Schaller$^{3}$,
\newauthor Richard Bower$^{3}$, Tom Theuns$^{3}$\\
$^{1}$ Leiden Observatory, Leiden University, P.O.\ Box 9513, NL-2300 RA Leiden, The Netherlands\\
$^{2}$ Astrophysics Research Institute, Liverpool John Moores University, 146 Brownlow Hill, Liverpool L3 5RF, UK\\
$^{3}$ Institute for Computational Cosmology, Department of Physics, University of Durham, South Road, Durham, DH1 3LE, UK }

\pubyear{2016}

\begin{document}
\label{firstpage}
\pagerange{\pageref{firstpage}--\pageref{lastpage}}
\maketitle

\begin{abstract}
We use the hydrodynamical EAGLE simulation to study the magnitude and origin of the scatter in the stellar mass - halo mass relation for
central galaxies. We separate cause and effect by correlating stellar masses in the baryonic simulation with halo properties in a matched dark matter only (DMO) simulation. The scatter in stellar mass increases with redshift and decreases with halo mass. At $z = 0.1$ it declines from 0.25 dex at $M_{200, \rm DMO} \approx 10^{11}$ M$_{\odot}$ to 0.12 dex at  $M_{200, \rm DMO} \approx 10^{13}$ M$_{\odot}$, but the trend is weak above $10^{12}$ M$_{\odot}$.  For $M_{200, \rm DMO} < 10^{12.5}$ M$_{\odot}$ up to 0.04 dex of the scatter is due to scatter in the halo concentration. At fixed halo mass, a larger stellar mass corresponds to a more concentrated halo. This is likely because higher concentrations imply earlier formation times and hence more time for accretion and star formation, and/or because feedback is less efficient in haloes with higher binding energies. The maximum circular velocity, $V_{\rm max, DMO}$, and binding energy are therefore more fundamental properties than halo mass, meaning that they are more accurate predictors of stellar mass, and we provide fitting formulae for their relations with stellar mass. However, concentration alone cannot explain the total scatter in the $M_{\rm star} - M_{200, \rm DMO}$ relation, and it does not explain the scatter in $M_{\rm star} -V_{\rm max, DMO}$. Halo spin, sphericity, triaxiality, substructure and environment are also not responsible for the remaining scatter, which thus could be due to more complex halo properties or non-linear/stochastic baryonic effects. \end{abstract} 




\section{Introduction}
The formation of structure in a universe consisting of dissipationless dark matter particles and dark energy is well understood and can be modelled with large $N$-body simulations, such that the halo mass function and the clustering of haloes can be predicted to high precision for a given set of cosmological parameters \citep[e.g.][]{Springel2006}. 

However, observations measure the masses and clustering of galaxies rather than dark matter haloes, so it is of utmost importance to connect stellar masses to dark matter halo masses. It is much more difficult for simulations to reproduce the observed stellar masses, as this requires a thorough understanding of the baryonic (feedback) processes involved, which are generally highly non-linear, complex and couple to a wide range of spatial scales. Therefore, a key goal of modern galaxy formation theory is to find the correlation or relation between the halo mass function and the stellar mass function. 

The relation between stellar mass and halo mass is related to the efficiency of star formation, and to the strength of feedback from star formation (e.g. radiation pressure from hot young stars, stellar winds or supernovae) and Active Galactic Nuclei (AGN, e.g. quasar driven outflows or heating due to radio jets that prevent gas from cooling). By matching the abundances of observed galaxies and simulated dark haloes ranked by stellar and dark matter mass respectively, we can infer that the relation is steeper for low-mass centrals than for high mass central galaxies \citep[e.g.][]{ValeOstriker2004,Kravtsov2014}. There is no tight relation between halo mass and stellar mass for satellite galaxies because of environmental processes such as tidal stripping, which is more efficient for the extended dark halo than for the stars. For the remainder of this paper, we therefore focus on central galaxies only. 

The evolution of galaxies is thought to be driven by the growth of halo mass \citep[e.g.][]{WhiteRees1978,Blumenthal1984}, as assumed by halo models and semi-analytical models (SAMs, e.g. \citealt{Henriques2015,Lacey2015}) and related techniques such as abundance matching \citep[e.g.][]{BerlindWeinberg2002,Yang2003,Behroozi2010,VdBosch2013}. However, both abundance matching models and observations suggest that there exists scatter in the stellar mass - halo mass (SMHM) relation  \citep{More2011,Moster2013, Behroozi2013,ZuMandelbaum2015}, meaning that halo masses alone cannot be used to predict accurate stellar masses. This could mean that there is also a second halo property which might explain (part of) the scatter in the stellar mass - halo mass (SMHM) relation, for example the formation time \citep[e.g.][]{Zentner2014}, or that there is a halo property other than mass which is more strongly correlated to stellar mass, such as the circular velocity \citep[e.g.][]{Conroy2006,Trujillo-Gomez2011}. 

\begin{figure}
\includegraphics[width=8.6cm]{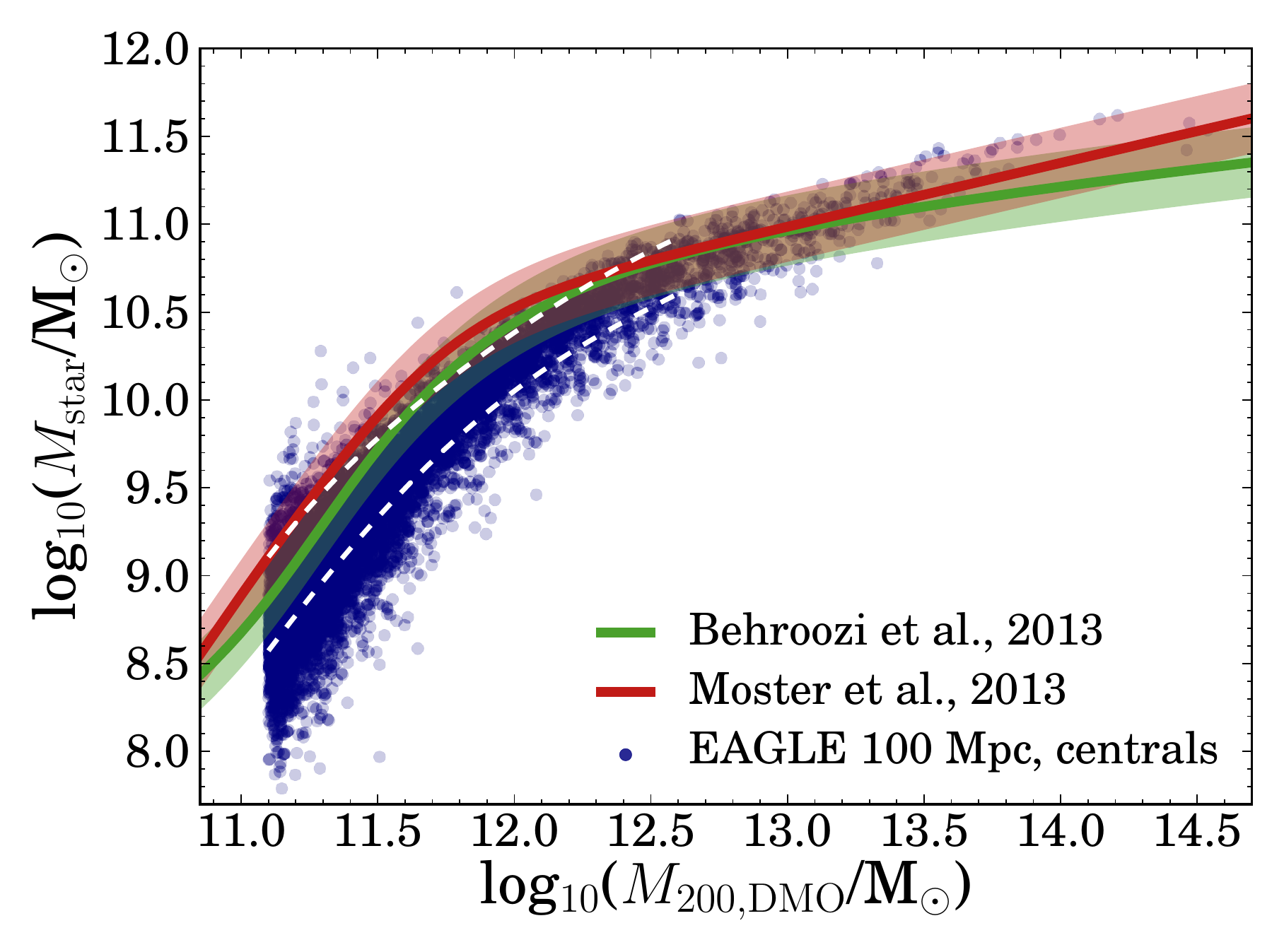}
\caption{\small{Relation between the stellar mass of central EAGLE galaxies and halo mass in the matched DMO simulation. The white dashed lines highlight the measured 1$\sigma$ scatter in the region where individual points are saturated. Also shown are results obtained from abundance matching to observations \citep{Behroozi2013,Moster2013}, including a shaded region indicating their 1$\sigma$ scatter. It can be seen that the slope changes at a halo mass around $10^{12}$ M$_{\odot}$, which is the mass at which the galaxy formation efficiency peaks. }}
\label{fig:smhm}
\end{figure}

In this paper, we use simulated galaxies from the EAGLE project \citep{Schaye2014,Crain2015} to assess which halo property can be used to predict stellar masses most accurately, and how it is related to the scatter in the stellar mass - halo mass relation, see Fig. $\ref{fig:smhm}$. EAGLE is a hydrodynamical simulation for which the feedback from star formation and AGN has been calibrated to reproduce the $z=0.1$ stellar mass function, galaxy sizes and the black hole mass - stellar mass relation. Because the simulation accurately reproduces many different observables and their evolution \citep[e.g.][]{Schaye2014,Furlong2015,Furlong2014,Trayford2016}, it is well suited for further studies of galaxy formation. 

The properties of dark matter haloes can be affected by baryonic processes \citep[e.g.][]{Bryan2013,Velliscig2014,Schaller2014profile}. For example, efficient cooling of baryons can increase halo concentrations. For our purposes, it is therefore critical to connect stellar masses to dark matter halo properties from a matched dark matter only simulation. Otherwise, it would be impossible to determine whether a given halo property is a cause or an effect of efficient galaxy formation. In order to find which halo property is most closely related to stellar mass, we thus use halo properties from the dark matter only version of EAGLE, which has the same initial conditions, box size and resolution as its hydrodynamical counterpart.

An important caveat in studying the scatter in a galaxy scaling relation in general is that many properties are correlated. For example, the scatter in the SMHM relation by construction can not correlate strongly with any property that correlates strongly with halo mass. This way, an actual physical correlation can be hidden. As many halo properties are related to halo mass \citep[e.g.][]{Jeeson-Daniel2011}, we should therefore be careful to only correlate the residuals of the SMHM relation to properties that are weakly or, ideally, not correlated with halo mass. We therefore use only dimensionless halo properties to study the origin of scatter in the SMHM relation.

This paper is organised as follows. The simulations and our analysis methods are presented in \S 2. In \S 3 we study which halo property is related most closely to stellar mass. We study the origin of scatter in the SMHM relation and the $M_{\rm star}-V_{\rm max, DMO}$ relation in \S 4. We show how we can predict more accurate stellar masses with a combination of halo properties in \S 5. In \S 6 we show the redshift evolution of the SMHM relation and its scatter. We discuss our results and compare with the literature in \S 7. Finally, \S 8 summarises the conclusions.

\section{Methods}
\subsection{The EAGLE simulation project}
In our analysis, we use central galaxies from the (100 cMpc)$^3$ reference {\sc EAGLE} model at redshift $z=0.101$, with a resolution such that a galaxy with a mass of $M_{\rm star} =10^{10}$ M$_{\odot}$ (such as the Milky Way) is sampled by $\sim 10,000$ star particles. The hydrodynamical equations are solved using the smoothed particle hydrodynamics (SPH) $N$-body code {\sc Gadget 3}, last described by \cite{Springel2005Gadget}, with modifications to the hydrodynamics solver \citep{Hopkins2013,DallaVecchia2016,SchallerSPH}, the time stepping \citep{Durier2012} and new sub-grid physics. There are $2 \times 1504^3$ particles with masses $1.8\times10^6$ M$_{\odot}$ (baryonic) and $9.7\times10^6$ M$_{\odot}$ (dark matter). The resolution has been chosen to resolve the Jeans scale in the warm (T$\sim10^4$ K) interstellar medium (at least marginally). {\sc EAGLE} uses a Planck cosmology \citep{PlanckParam}. The halo and galaxy catalogues and merger trees from the {\sc EAGLE} simulation are publicly available \citep{McAlpine2015}.

\begin{table}
\begin{center}
\begin{small}
\caption{\small{The properties of the simulated galaxies and haloes that are considered in our analysis. The stellar mass is from the reference EAGLE model, while the other properties are from the matched dark matter-only simulation. See \S 2.3 for detailed definitions of properties. }} 
\begin{tabular}{ ll }
$\bf Property$ & $\bf Description $  \\ \hline
Dimensional & \\
$M_{\rm star}$   & Stellar mass inside 30 kpc, in M$_{\odot}$ \\ 
$M_{200, \rm DMO}$ & Mass, in M$_{\odot}$ \\
$M_{\rm core, DMO}$ & Mass within NFW scale radius, in M$_{\odot}$ \\
  $\sigma_{2500, \rm DMO}$  & Central velocity dispersion, in km/s \\
  $V_{\rm max, DMO}$ & Maximum circular velocity, in km/s \\
  $E_{2500, \rm DMO}$ & Binding energy, in M$_{\odot}$ km$^{2}$/s$^{2}$\\ 
 $V_{\rm peak, DMO}$ & Highest $V_{\rm max}$ in a galaxy's history, in km/s\\
 $V_{\rm relax, DMO}$ & Highest $V_{\rm max}$ while halo was relaxed, in km/s\\ \hline

  Dimensionless  & \\
$N_{\rm 2Mpc, DMO}$  & Total number of subhalos within 2 Mpc \\
$N_{\rm 10Mpc, DMO}$  & Total number of subhalos within 10 Mpc \\
$c_{200, \rm DMO}$  & Concentration \\
$\lambda_{200, \rm DMO}$ & Spin \\
 $s_{\rm DMO}$  & Sphericity\\
 $T_{\rm DMO}$ & Triaxiality \\
  substructure  & Mass fraction in bound substructures in a halo\\
  $z_{0.5, \rm DMO}$  & Assembly redshift \\
  \hline
\end{tabular}
\label{tab:properties}
\end{small}
\end{center}
\end{table}
For hydrodynamical simulations of galaxy formation, the implementation of sub-grid physics is critical \citep[e.g.][]{Schaye2010,Scannapieco2012}. The included sub-grid models account for radiative cooling by the eleven most important elements \citep{Wiersma2009cooling}, star formation \citep{SchayeVecchia2008} and chemical enrichment \citep{Wiersma2009enrich}, feedback from star formation \citep{VecchiaSchaye2012}, growth of black holes \citep{Springel2005Nat,Rosas2015,Schaye2014} and feedback by AGN \citep{BoothSchaye2009}. Galactic winds develop naturally without predetermined mass loading factors, velocities or directions, without any explicit dependence on dark matter properties and without disabling the hydrodynamics or the radiative cooling. This is achieved by injecting the feedback energy thermally using the stochastic implementation of \cite{DallaVecchia2012}, which reduce numerical radiative losses. 
As discussed by \cite{Crain2015}, the $z\approx0$ galaxy stellar mass function can be reproduced even without tuning the feedback parameters. However, the feedback needs to be calibrated in order to simultaneously reproduce present-day galaxy sizes, which in turn leads to agreement with many other galaxy scaling relations.

\subsection{Halo definition and matching between simulations}
Haloes and galaxies are identified using the two step Friends-of-Friends (FoF) \citep[e.g.][]{Einasto1984} and {\sc subfind} \citep{Springel2001,Dolag2008} algorithms. First, the FoF-algorithm groups particles together using a linking length of 0.2 times the mean inter-particle distance \citep{Davis1985}. Then,  {\sc subfind} identifies subhalos as local overdensities whose membership is defined by the saddle points in the density distribution. The particles are then verified to be gravitationally bound to the substructure. The central galaxy is the subhalo at the minimum potential of the FoF-group. Following \cite{Schaye2014}, we use a spherical 30 proper kpc aperture, centred on the central subhalo in each FoF-group, to measure the stellar masses of each central galaxy. 

Dark matter halo properties are taken from the dark matter only version of {\sc EAGLE} (DMO), which has the same initial conditions (phases and amplitudes of initial Gaussian field) and resolution as the reference model. Haloes in the DMO and {\sc EAGLE} reference simulations were matched as described by \cite{Schaller2014}. In short, the 50 most-bound dark matter particles were selected for each halo in the reference model. These particles were located in the DMO model and haloes were matched if at least 25 of these particles belong to a single FOF halo in the DMO simulation. We note that for the halo masses discussed here (for the sample selection see \S 2.5), $>99$\% of the haloes are matched successfully.

\subsection{Definitions of halo-properties}
 We study two classes of (dark matter) halo properties: dimensional and dimensionless. An overview of the properties, which are defined in this section, is given in Table $\ref{tab:properties}$.
 
 \subsubsection{Dimensional halo properties}
In addition to stellar mass and halo mass ($M_{200, \rm DMO}$), dimensional properties that we consider are the core mass ($M_{\rm core, DMO}$), the maximum circular velocity at $z=0.1$ ($V_{\rm max}$) and in the halo's history ($V_{\rm peak}$), the central velocity dispersion ($\sigma_{\rm 2500, DMO}$) and the halo binding energy ($E_{2500,\rm DMO}$). While our main focus is on the stellar mass - halo mass relation, we use the other dimensional halo properties to investigate which halo property correlates best with stellar mass. Note that we vary our definition of stellar mass in Appendix A1. 

$M_{200, \rm DMO}$ is used as the halo mass, which is the total mass contained within $R_{200, \rm DMO}$, the radius within which the enclosed over-density is 200 times the critical density. We study the effect of changing the definition to 500 and 2500 times the critical density in \S 3. $V_{\rm max}$ is the maximum circular velocity, ${\rm max} ( \sqrt{\frac{GM(<R)}{R}})$. $V_{\rm peak}$ is the maximum circular velocity a halo had over its history (for central galaxies this is typically similar to the current $V_{\rm max}$, as shown for EAGLE by \citealt{Chaves2015}). We also include $V_{\rm relax}$, the maximum circular velocity of a halo during the part of its history when the halo was relaxed, which correlates most strongly with stellar mass \cite{Chaves2015}\footnote{Note that \cite{Chaves2015} included satellites, whereas we only consider central galaxies.}. In this definition, a halo is relaxed when the formation time is longer than the crossing time \citep[e.g.][]{Ludlow2012}. The formation time is defined as the time at which a fraction of $3/4$ of the halo mass was first assembled in the main progenitor (although using a fraction of $1/2$ leads to similar results, see \citealt{Chaves2015}).

Another definition of the halo mass is the halo core mass ($M_{\rm core, DMO}$), which is the mass inside the scale radius ($r_s$) of the NFW profile \citep[e.g.][]{Huss1999}. As highlighted by \cite{Diemer2013}, the evolution of $M_{200, \rm DMO}$ can be split into two stages: an initial growth of mass inside the $z=0$ scale radius (growth of the core mass, e.g. \citealt{Ludlow2013,Correa2015}), followed by ``pseudo-evolution'' due to the decreasing critical density of the Universe with cosmic time, during which the core mass remains nearly constant. We compute the core mass using the NFW fits of \cite{Schaller2014} to obtain the scale radius. Typically, the core mass is $\approx0.15\times M_{200, \rm DMO}$, although there is significant scatter of 0.2 dex.

The halo binding energy is related to the halo mass and concentration. Galaxy formation may be more efficient in a halo with a higher binding energy \citep[e.g.][]{BoothSchaye2010,BoothSchaye2011}, since it will be harder for stellar and black hole feedback to drive galactic winds out of the galaxy. We compute the binding energy at three different radii: $R_{200, \rm DMO}$, $R_{500, \rm DMO}$ and $R_{2500, \rm DMO}$, using $E_{200, \rm DMO} = M_{200, \rm DMO} \, \sigma_{\rm 200, DMO}^2$, where $\sigma_{\rm 200, DMO}$ is the velocity dispersion within $R_{200, \rm DMO}$ (and similarly for $R_{500, \rm DMO}$ and $R_{2500, \rm DMO}$). As we are generally interested in stellar mass, which is concentrated in the centres of haloes, we focus on the binding energy and velocity dispersion of dark matter particles within $R_{2500, \rm DMO}$. This radius ranges from $R_{2500, \rm DMO}\approx 50$ kpc for $M_{\rm star} = 10^{9.5}$ M$_{\odot}$ to $R_{2500, \rm DMO} \approx 350$ kpc for galaxies with $M_{\rm star} = 10^{11.5}$ M$_{\odot}$. $R_{2500, \rm DMO}$ is typically $\approx0.3\times R_{200, \rm DMO}$, and typically $\approx2\times r_{s}$, where $r_{s}$ is the NFW scale radius.

\subsubsection{Dimensionless halo properties} 
Dimensionless halo properties are generally related to the shape of the halo (such as triaxiality, sphericity, concentration and substructure), the environment (such as the number of neighbours) or its spin. These dimensionless properties are considered when we study the scatter in scaling relations. 

The halo concentration was obtained by fitting a Navarro-Frenk-White (NFW) profile \citep{Navarro1997,NFW1997} to the dark matter particles in the halo, as described by \cite{Schaller2014}. The concentration is defined as $c_{200, \rm DMO}=R_{200, \rm DMO}/r_s$. 

The dimensionless spin parameter, $\lambda_{200, \rm DMO}$, is defined as in \cite{Bullock2001}, $\lambda_{200, \rm DMO} = \frac{j}{\sqrt{2}V_{200, \rm DMO}R_{200, \rm DMO}}$, where $j=L/M$ is the specific angular momentum.

We quantify the shape of the halo with the sphericity, $s$, and triaxiality, $T$, parameters. The sphericity is defined as $s = c/a$, where $c$ and $a$ are the minor and major axes of the inertia tensor \citep[e.g.][]{Bett2007}. The halo triaxiality is defined as  $T = \frac{a^2-b^2}{a^2-c^2}$ \citep{Franx1991}.

The environment of the halo, $N_{X\, \rm Mpc}$, is quantified by the number of neighbours within a distance of $X$ Mpc. The number of neighbours, defined as the number of subhalos (including satellites) with a total dark matter mass above $10^{10}$M$_{\odot}$, is measured within spheres of 2 Mpc ($N_{\rm 2Mpc}$) and 10 Mpc ($N_{\rm 10Mpc}$). 

The substructure parameter quantifies the environment of the central galaxy within the halo. It is defined as the fraction of the total mass of a FoF halo in bound substructures with dark matter mass above $10^{10}$M$_{\odot}$.

The assembly history of a halo is quantified by $z_{0.5,\rm DMO}$, the redshift at which half of the halo mass has been assembled into a single progenitor subhalo. We use the EAGLE merger trees \citep{McAlpine2015} to track the dark matter mass of the haloes from $z=4$. For a halo at a fixed redshift, we select all the progenitors in the previous snapshot. The mass of the halo at that previous redshift is then the halo mass of its most massive progenitor. We thus obtain a mass assembly history for each halo and measure the formation redshift using a spline interpolation of the masses at the different snapshots.\footnote{The snapshot redshifts are $z=[0.10$, 0.18, 0.27, 0.37, 0.50, 0.62, 0.74, 0.87, 1.00, 1.26, 1.49, 1.74, 2.01, 2.28, 2.48, 3.02, 3.53, 3.98].}

\begin{figure}
\includegraphics[width=8.6cm]{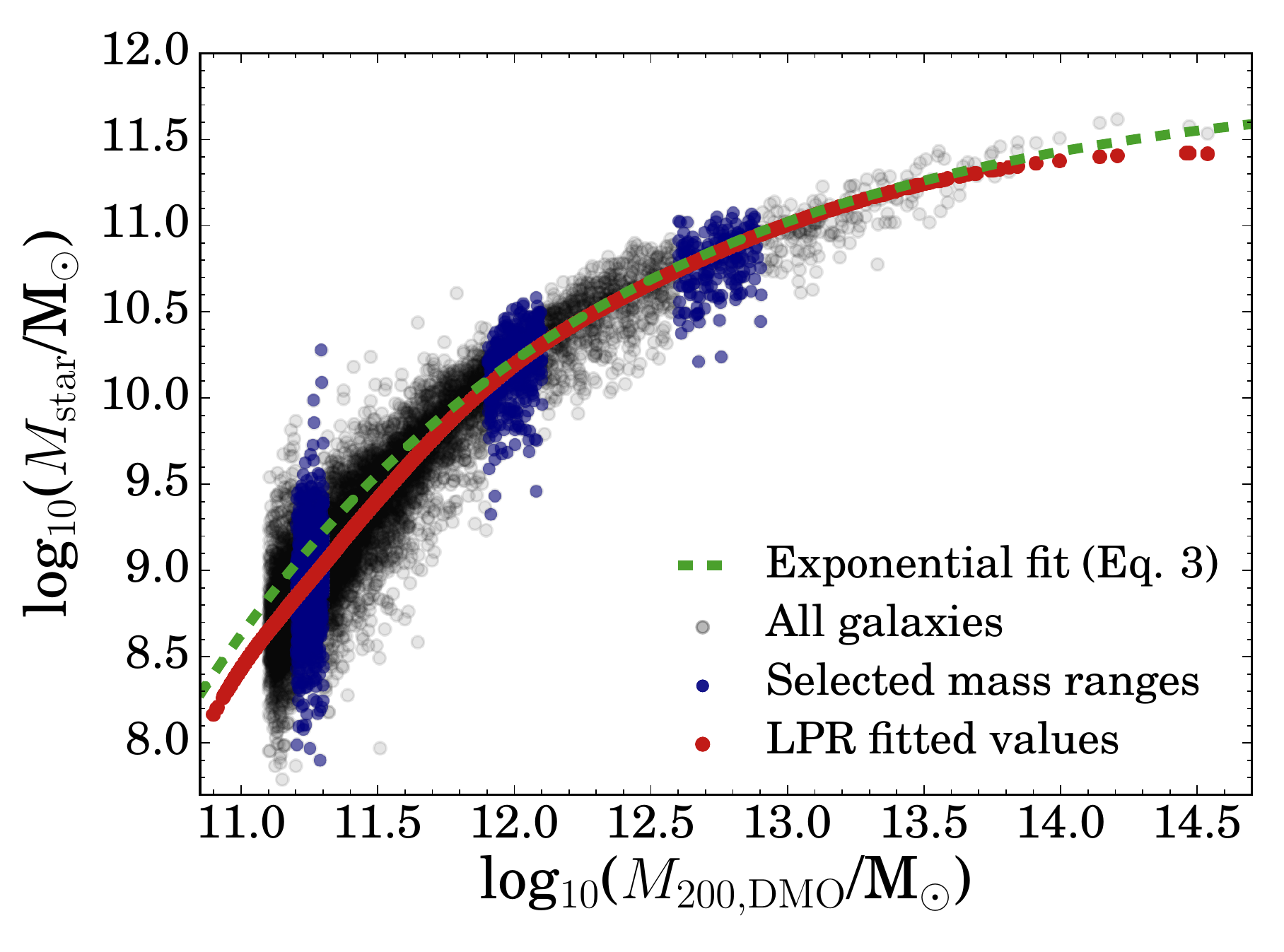}
\caption{\small{Relation between stellar mass in the EAGLE simulation and halo mass in the matched DMO simulation, illustrating the method to obtain residuals. The red points show the relation fitted using the non-parametric LPR method, see \S 2.4.1. The green line is the exponential fit specified by Eq. $\ref{eq2}$. The points marked in blue correspond to the three mass regimes mentioned throughout the text. Black points are galaxies not included in our analysis, but included in the LPR estimate of the relation.}}
\label{fig:smhm_res}
\end{figure} 

\subsection{Obtaining residuals of scaling relations}
We quantify the scatter in scaling relations as the 1$\sigma$ vertical offset from the mean relation (the residual). The mean relations are estimated in two ways: using non-parametric and parametric methods. The benefit of non-parametric methods is that they do not require an assumed functional form, but the downside is that they are less easily reproducible and perform less well at the limits of the dynamic range. 

\subsubsection{Non-parametric method: local polynomial regression}
For the non-parametric approach, we use the local polynomial regression method (LPR, also known as locally-weighted scatterplot smoothing, LOWESS; \citealt{Cleveland1979}). In short, for each data point $X_i=[x_i,y_i]$ a fitted value $f_i$ is obtained using a local linear fit (described below), for which only the nearest half of the other data points is used. Our results are insensitive to changes in the fraction between 0.3 and 0.6 of the data that is used, except for the highest masses where a larger fraction results in an underestimate of the relation, or the lowest masses where smaller fractions result in greater noise. The following weight is then applied to each of the closest half of the data points:
\begin{equation}
\label{eq1}
w_{ij} = (1 - (\frac{d}{{\rm max}(d)})^3)^3,
\end{equation}
where $d=\sqrt{(x_j-x_i)^2+(y_j-y_i)^2}$ is the two-dimensional distance between points $X_i$ and $X_j$. Finally, a linear relation is fitted to each selected data point using the least squares method:
\begin{equation}
f_i = \sum\limits_j w_{ij} y_j / \sum\limits_j w_{ij}.
\end{equation}
From this linear relation, the fitted value, $f_i$, for $X_i$ is obtained. This procedure is repeated for each point. This method is included in the R statistical language\footnote{https://www.r-project.org} by B. D. Ripley, and the reference for more information is \cite{LPRBOOK}. The main benefit of this method is that it can handle non-trivial relations without assuming a functional form. 

For the LPR procedure, we include all galaxies with a halo mass greater than 10$^{11}$ M$_{\odot}$ in the matched DMO simulation (corresponding to stellar components with $\gtrsim 500$ star particles). The resulting values of $f_i$ are shown using red symbols in Fig. $\ref{fig:smhm_res}$. Note that there are as many red points as grey points, but that the red points appear as a line where they are close to each other. After this procedure, we find that the scatter, the 1$\sigma$ standard deviation of the residuals ($y_i - f_i$; $\sigma$($\Delta$log$_{10}$ $M_{\rm star}$($M_{200, \rm DMO}$))), ranges from 0.15 dex to 0.27 dex, depending on the halo mass (see e.g. Table $\ref{tab:bins}$ and Fig. $\ref{fig:scatter_mstar_haloproperties}$). A shortening of this method is that its accuracy depends on the number density of neighbouring points in the two dimensional plane. Therefore, it is less accurate at the highest masses ($M_{200, \rm DMO} > 10^{13.5}$ M$_{\odot}$, see Fig. $\ref{fig:smhm_res}$) where there are fewer points and the available neighbours are strongly biased toward lower masses. Haloes of these masses are however not included in our analysis because of their small number in the simulation. 
\begin{table}
\begin{center}
\begin{small}
\caption{\small{Fitted parameters for relations between stellar mass and the listed DMO halo properties using the functional form from Eq. 2.}}
\begin{tabular}{lrrrr }
 Halo property & $\alpha$ &$\beta$  & $\gamma$ & $\chi^2_{\rm red}$\\ \hline
$M_{200, \rm DMO}$ & $11.85^{+0.32}_{-0.18}$ &$-0.68^{+0.13}_{-0.12}$ &$8.65^{+1.17}_{-1.32}$ & 0.12\\
$V_{\rm max, DMO}$ & $11.56^{+0.62}_{-0.29}$ & $-2.67^{+0.73}_{-0.76}$ &$6.28^{+1.35}_{-1.35}$ & 0.08\\
$V_{\rm peak, DMO}$ & $11.66^{+0.57}_{-0.32}$ & $-2.46^{+0.62}_{-0.85}$ &$5.92^{+1.53}_{-1.17}$ & 0.10\\
$V_{\rm relax, DMO}$ & $11.62^{+0.59}_{-0.28}$ & $-2.63^{+0.71}_{-0.76}$ &$6.20^{+1.39}_{-1.30}$ & 0.11\\
$E_{2500,\rm DMO}$ & $11.54^{+0.48}_{-0.17}$ & $-0.54^{+0.11}_{-0.03}$ &$8.77^{+0.36}_{-1.39}$ & 0.08\\
$\sigma_{2500, \rm DMO}$ & $11.49^{+0.66}_{-0.32}$ & $-2.77^{+0.78}_{-0.91}$ &$5.94^{+1.52}_{-1.22}$ & 0.04\\ \hline
\end{tabular}
\label{tab:fits}
\end{small}
\end{center}
\end{table}

\subsubsection{Parametric method: functional fit}
In addition to the non-parametric LPR fits, we perform parametric fits to the relations between stellar mass and dark matter halo properties. We use the following functional form in log-log space, which has three free parameters ($\alpha$, $\beta$, $\gamma$):
\begin{equation}
\label{eq2}
{\rm log}_{10} (M_{\rm star}/{\rm M}_{\odot}) = \alpha - e^{\beta\, {\rm log}_{10}(M_{200, \rm DMO}/\rm M_{\odot})+\gamma}.
\end{equation}
In this equation, the halo property used is the halo mass ($M_{200, \rm DMO}$), but it can be replaced by the other properties from \S 2.3.1. Because our sample of galaxies is dominated by the lowest-mass galaxies, we weight our fit, such that galaxies at all masses contribute equally. To do this, we compute the average stellar mass in halo mass bins of 0.1 dex and compute the standard deviation of the stellar masses in each bin. We only include bins that contain more than ten haloes (so up to $M_{200, \rm DMO} \approx10^{13.5}$ M$_{\odot}$). Using these bins and using the standard deviations as errors, we fit Eq. $\ref{eq2}$ by minimising the $\chi^2$ value. We start with a large, but sparse, three dimensional grid of allowed values for the three parameters. After a first estimate of the values, we increase the resolution in a smaller range of allowed values to obtain our best-fit values. For the SMHM relation, we find best-fitting values of $\alpha = 11.85^{+0.32}_{-0.18}$, $\beta = -0.68^{+0.13}_{-0.12}$ and $\gamma = 8.65^{+1.17}_{-1.32}$.\footnote{We note that Eq. $\ref{eq2}$ can alternatively also be written as: log$_{10}(\frac{M_{\rm star}}{\rm M_{\odot}}$) = $\alpha - e^{\gamma} (\frac{M_{200, \rm DMO}}{\rm M_{\odot}})^{\beta {\rm log}_{10} (e)}$. \\In this case, our best fit can be written as: \\${\rm log}_{10}(\frac{M_{\rm star}}{10^{10} {\rm M}_{\odot}}$) = $1.85-1.63 (\frac{M_{200, \rm DMO}}{10^{12} {\rm M}_{\odot}})^{-0.30}$.} The fit has a reduced $\chi^2$ of 0.12 for 27 degrees of freedom. It can be seen in Fig. $\ref{fig:smhm_res}$ that the parametric fit (green line) resembles the LPR values (red points) very well, except for the highest and lowest masses, for which the LPR method is less successful. This is because the LPR method is slightly biased towards the edges of parameter space, which can be overcome when the number density is sufficiently large to include a smaller fraction in the fit a larger number density without adding noise.

\begin{figure*}
\begin{tabular}{cc}
\includegraphics[width=8.6cm]{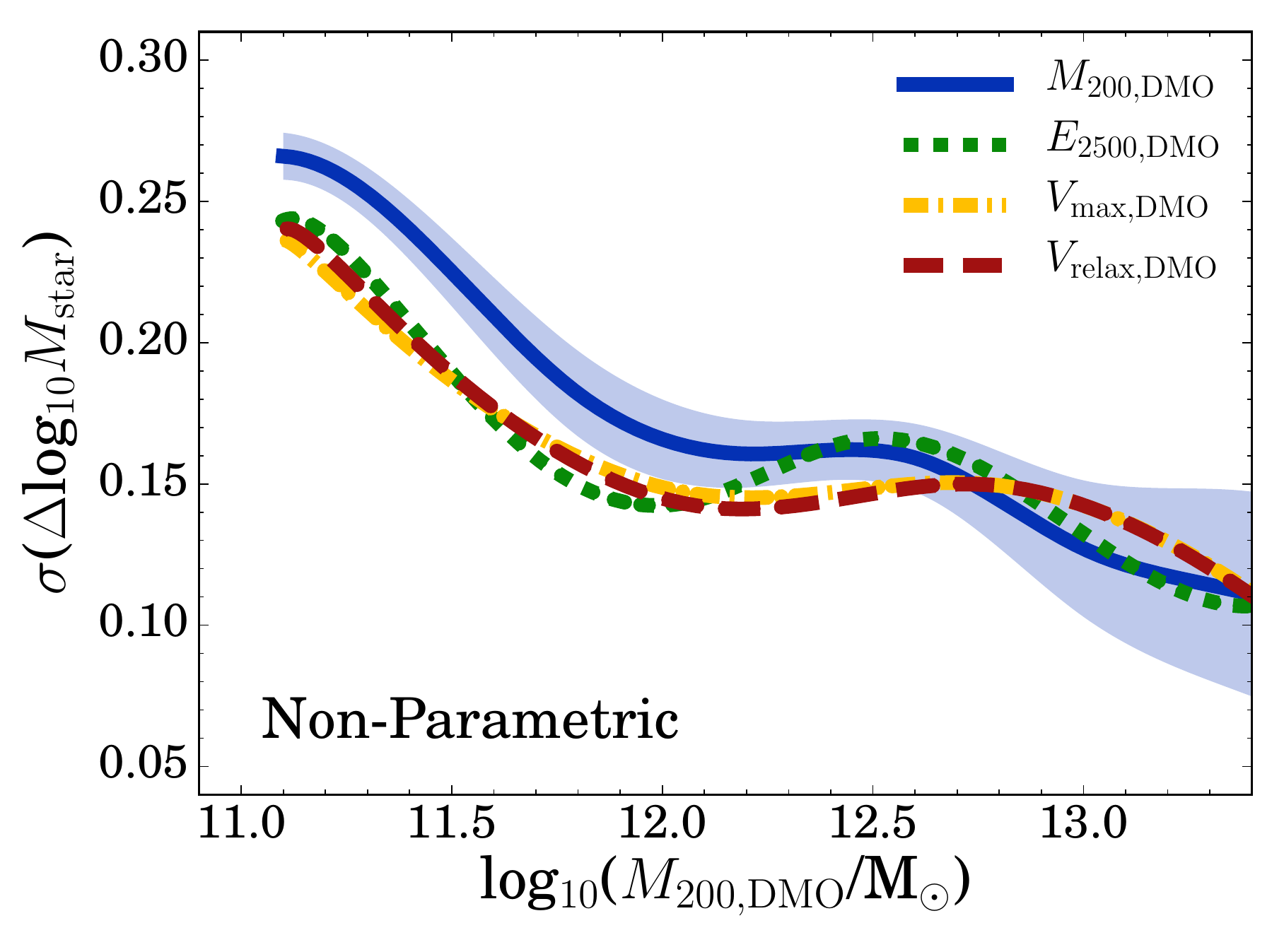} &
\includegraphics[width=8.6cm]{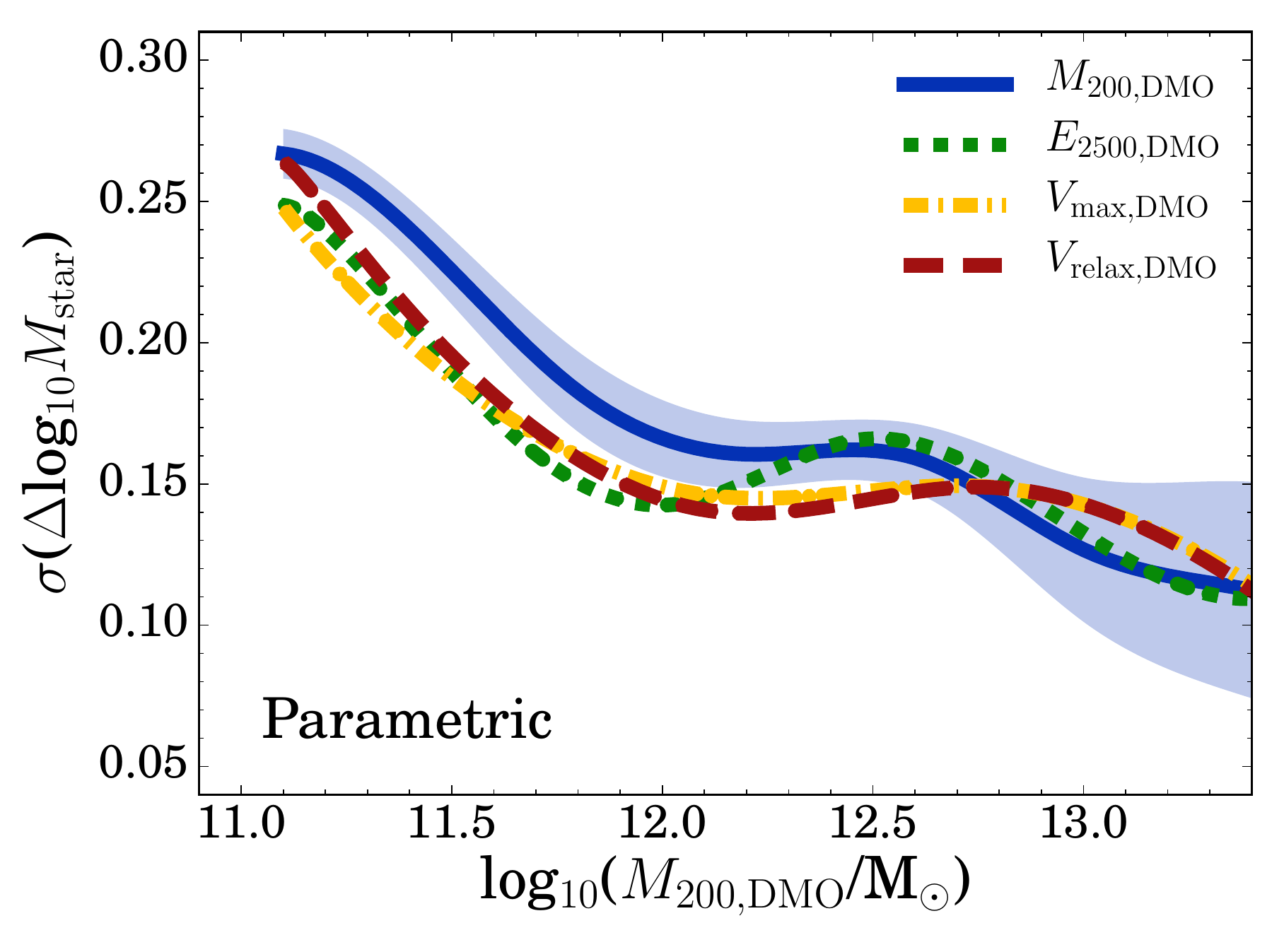} \\

\end{tabular}
\caption{\small{Scatter in the difference between true stellar masses (in the baryonic simulation) and stellar masses computed from the non-parametric ({\it left}) and parametric ({\it right}) fits to the relation between stellar mass and the different dark matter halo properties from the matched DMO simulation listed in the legend, as a function of DMO halo mass. We show the jackknife estimates of the errors in $\sigma$(log$_{10}$ $M_{\rm star}$($M_{200, \rm DMO}$)) as a function of $M_{200, \rm DMO}$ as a blue shaded region. The errors on $\sigma(\Delta{\rm log}_{10} M_{\rm star}$) for the other halo properties are similar. We note that for halo properties other than $M_{200, \rm DMO}$ we have binned $\sigma(\Delta{\rm log}_{10} M_{\rm star}$) in bins of the specific halo property, and plot the result as a a function of the corresponding $M_{200, \rm DMO}$ of that bin. In general, there is more scatter in the stellar mass - halo property relation at small stellar mass than at high stellar mass, irrespective of the halo property used. For $M_{200, \rm DMO}<10^{12.5}$ M$_{\odot}$, $M_{200, \rm DMO}$ is a less accurate predictor of stellar mass than $V_{\rm max, DMO}$, $V_{\rm relax, DMO}$ and $E_{2500,\rm DMO}$. }}
\label{fig:scatter_mstar_haloproperties} 
\end{figure*}

We also fit Eq. $\ref{eq2}$ to the relation between stellar mass and $E_{2500,\rm DMO}$, $V_{\rm max, DMO}$, $V_{\rm peak, DMO}$, $V_{\rm relax, DMO}$ and $\sigma_{2500, \rm DMO}$ with the same method as described above. The results are summarised in Table $\ref{tab:fits}$. Using these equations, the dispersion in the residuals of the SMHM relation, $\sigma$($\Delta$log$_{10}$ $M_{\rm star}$($M_{200, \rm DMO}$)), is 0.15-0.26 dex, depending on the stellar mass range (see Fig. $\ref{fig:scatter_mstar_haloproperties}$).
 
For infinitesimally small bins of halo mass, the dispersion in the residuals of the SMHM relation is equal to the scatter in stellar mass at fixed halo mass, $\sigma$(log$_{10}$ $M_{\rm star}$($M_{200, \rm DMO}$)), and we will now therefore abbreviate this to $\sigma(\Delta{\rm log}_{10} M_{\rm star}$) for simplicity in the remainder in the text.

\begin{table}
\begin{center}
\begin{small}
\caption{\small{Properties of the three halo mass samples. Different columns show different DMO halo mass ranges, the average stellar mass and the 1$\sigma$ dispersion of the residuals of the stellar mass - halo mass relation, abbreviated as $\sigma(\Delta{\rm log}_{10} M_{\rm star}$) .}}
\begin{tabular}{ c|l|c }
Halo mass range   &$ <M_{\rm star} > $ & $\sigma(\Delta{\rm log}_{10} M_{\rm star}$) \\ 
 (M$_{\odot}$) & (M$_{\odot}$) & (dex) \\ \hline
 11.2 $<$ log$_{10}$($M_{200, \rm DMO}$) $<$ 11.3  & $8.7\times10^8$  & 0.26\\
  11.9 $<$ log$_{10}$($M_{200, \rm DMO}$) $<$ 12.1  & $1.7\times10^{10}$  & 0.16 \\
  12.6 $<$ log$_{10}$($M_{200, \rm DMO}$) $<$ 12.9  & $6.5\times10^{10}$ & 0.16 \\ \hline
\end{tabular}
\label{tab:bins}
\end{small}
\end{center}
\end{table}

\subsection{Sample selection and mass range dependence}
We initially select all central galaxies at $z=0.1$ with a halo mass of $M_{200} > 10^{11}$ M$_{\odot}$ in the EAGLE simulation (and use these for fitting). However, due to small differences between $M_{200}$ and $M_{200, \rm DMO}$ we restrict our analysis to galaxies with $M_{200, \rm DMO} > 10^{11.1}$ M$_{\odot}$ to avoid any biases which could arise from the influence of baryons on the dark matter halo mass of the lowest halo masses. 

In order to estimate the scatter in the SMHM relation as a function of halo mass, we perform the following steps: for each halo, we first obtain the residual relative to the main relation between stellar mass and the halo property (using either the non-parametric or parametric method). We then divide our sample of galaxies in bins (with width 0.4 dex for bins of halo mass, 0.6 dex for bins of $E_{2500,\rm DMO}$ and 0.2 dex for bins of $V_{\rm max}$), and compute the 1$\sigma$ dispersion in the residual values of galaxies in each bin. We interpolate the values of the 1$\sigma$ scatter as a function of halo mass and show this for the different halo properties in Fig. $\ref{fig:scatter_mstar_haloproperties}$. Errors on $\sigma(\Delta{\rm log}_{10} M_{\rm star}$) are estimated using the jackknife method. This means that we split the simulated volume in eight sub-domains of (50 cMpc)$^3$ and compute the 1$\sigma$ spread of residuals of the SMHM relation of the galaxies in each sub-box (for each bin of halo mass). Errors become significantly large at $M_{200, \rm DMO} \gtrsim 10^{12.7}$ M$_{\odot}$ because of the limited number of massive haloes in the simulation.

As the correlations between halo properties and stellar mass might depend on the mass range, we also investigate how correlations between residuals and halo properties vary with halo mass (or circular velocity or binding energy, depending on the relevant halo property). We therefore compare galaxies in three narrow ranges of halo mass throughout the text. These intervals are listed in Table $\ref{tab:bins}$ and they are illustrated as blue points in Fig. $\ref{fig:smhm_res}$. The lowest halo mass range is typical of dwarf galaxies, the middle of Milky Way like galaxies and the highest mass range of massive galaxies (the number of galaxies in a fixed range of halo masses declines quickly with mass, such that our bin widths increase with mass).

\section{Correlations between stellar mass and DMO halo properties} 
In this section we explore which halo property correlates best with the stellar mass of central galaxies and is therefore the most fundamental. 

\begin{table}
\begin{center}
\caption{\small{Amount of scatter in stellar mass over all masses, as defined by the 1$\sigma$ spread in the residuals from the non-parametric relation between stellar mass and the relevant DMO halo property. The column on the right shows the Spearman correlation rank coefficient for the relation between stellar mass and the halo property. }}
\begin{tabular}{ l|r|r|r }
Halo-property& 1$\sigma$ scatter with $M_{\rm star}$ & R$_s$  \\ \hline
$M_{200, \rm DMO}$ & 0.24 & 0.92 \\
$M_{500, \rm DMO}$ & 0.22 & 0.93 \\
$M_{2500, \rm DMO}$ & 0.21 & 0.93 \\
$M_{\rm core, DMO}$ & 0.33 & 0.85 \\
$M_{\rm 200, mean, DMO}$ & 0.24 & 0.91 \\
$E_{200, \rm DMO}$ & 0.23 & 0.92 \\
$E_{500, \rm DMO}$ & 0.22 & 0.93 \\
$E_{2500,\rm DMO}$ & 0.21 & 0.93 \\
$\sigma_{200, \rm DMO}$ & 0.25 & 0.91 \\
$\sigma_{500, \rm DMO}$ & 0.24 & 0.91 \\
$\sigma_{2500, \rm DMO}$ & 0.24 & 0.92 \\
$V_{\rm max, \rm DMO}$ & 0.21 & 0.93 \\
$V_{\rm peak, \rm DMO}$ & 0.24 & 0.93 \\
$V_{\rm relax, \rm DMO}$ & 0.21 & 0.93 \\ \hline
\end{tabular}
\label{tab:spearmanrank_analysis}
\end{center}
\end{table}

In order to determine which halo property correlates most strongly with stellar mass, we perform a Spearman rank correlation (R$_s$) analysis. In a Spearman rank analysis, the absence of a relation between two properties results in R$_s =0$ and a perfect (anti-)correlation results in R$_s= (-)1$. We will call a correlation `strong' if |R$_s|>0.3$. For this value, a correlation of 70 data points is statistically significant at 99\% confidence. For our highest halo mass bin, consisting of 228 galaxies, a 99\% confidence significance is obtained for R$_s=0.17$ and higher.

We find that all dimensional halo properties are strongly correlated with stellar mass, with Spearman coefficients R$_s > 0.85$, see Table $\ref{tab:spearmanrank_analysis}$. The highest Spearman coefficients are found for $V_{\rm max, DMO}$, $V_{\rm peak, DMO}$, $V_{\rm relax, DMO}$ and the halo mass and binding energy at $R_{2500,\rm DMO}$ and $R_{500, \rm DMO}$, which all give R$_s = 0.93$. This indicates that the central binding energy or maximum circular velocity are the most fundamental halo properties, although the differences are marginal. 

Another way to study which halo property is the most fundamental, is by exploring how accurately a halo property can predict stellar masses, as a function of halo mass. By ``accuracy" we mean the 1$\sigma$ scatter in the difference between the predicted and true stellar masses, $\sigma(\Delta{\rm log}_{10} M_{\rm star}$). Predicted stellar masses are obtained with both the non-parametric and the parametric relations between stellar mass and halo properties (see \S 2.4), and the true stellar masses are those measured in the baryonic simulation. The number density-weighted averaged results are listed in Table $\ref{tab:spearmanrank_analysis}$. The scatter is largest for the core mass (0.33 dex) and smallest (0.21 dex) for the halo mass measured at $R_{2500, \rm DMO}$, $E_{2500,\rm DMO}$, $V_{\rm max, DMO}$ and $V_{\rm relax, DMO}$.
In Fig. $\ref{fig:scatter_mstar_haloproperties}$ we show the mass dependence of the results for the halo properties with the least scatter in the difference between predicted and true stellar masses. Note that we vary the definitions of halo mass in Fig. $\ref{fig:mhalo_variations}$ and of stellar mass in Appendix A. 

Regardless of the halo property or fitting method, we find that $\sigma(\Delta\rm log_{10} M_{\rm star}$) decreases from $\gtrsim 0.25$ dex at $M_{200, \rm DMO} \approx 10^{11.2}$ M$_{\odot}$ to $\gtrsim 0.15$ dex at $M_{200, \rm DMO} \approx 10^{12.2}$ M$_{\odot}$. We show in Appendix B that this is not an effect of the limited simulation volume. This is in contrast with the typical assumptions in halo models, which use a mass-independent scatter of $\sim0.20$ dex \citep[i.e.][]{Moster2013,Uitert2016}. Above $M_{200, \rm DMO} \gtrsim 10^{12.2}$ M$_{\odot}$, the uncertainties in $\sigma(\Delta\rm log_{10}M_{\rm star}$) are large enough (likely due to the limited simulation volume) that a constant scatter cannot be ruled out. Therefore, for the highest halo masses the decrease in the scatter with halo mass needs to be confirmed with larger simulation volumes.

We find that $V_{\rm max, DMO}$, $V_{\rm relax, DMO}$ and $E_{2500,\rm DMO}$ give similarly small $\sigma(\Delta{\rm log}_{10} M_{\rm star}$), while $M_{200, \rm DMO}$ performs somewhat worse for $M_{200, \rm DMO} < 10^{12.5}$M$_{\odot}$. However, using the parametric method, the differences are slightly smaller. This might mean that the chosen functional form is not optimal for $V_{\rm max}$, $V_{\rm relax}$ and $E_{2500,\rm DMO}$ at low masses. Perhaps more striking is the fact that regardless of the halo property, there is at least 0.15 dex scatter in stellar masses at $M_{\rm 200} < 10^{12}$M$_{\odot}$, indicating that processes other than those captured by our halo properties are important. Another feature is that the slope changes at a mass of $\approx10^{12}$ M$_{\odot}$, which coincides with the halo mass at which the galaxy bimodality arises and where feedback from AGN starts to become important \citep[e.g.][]{Bower2016}.

\begin{figure}
\includegraphics[width=8.6cm]{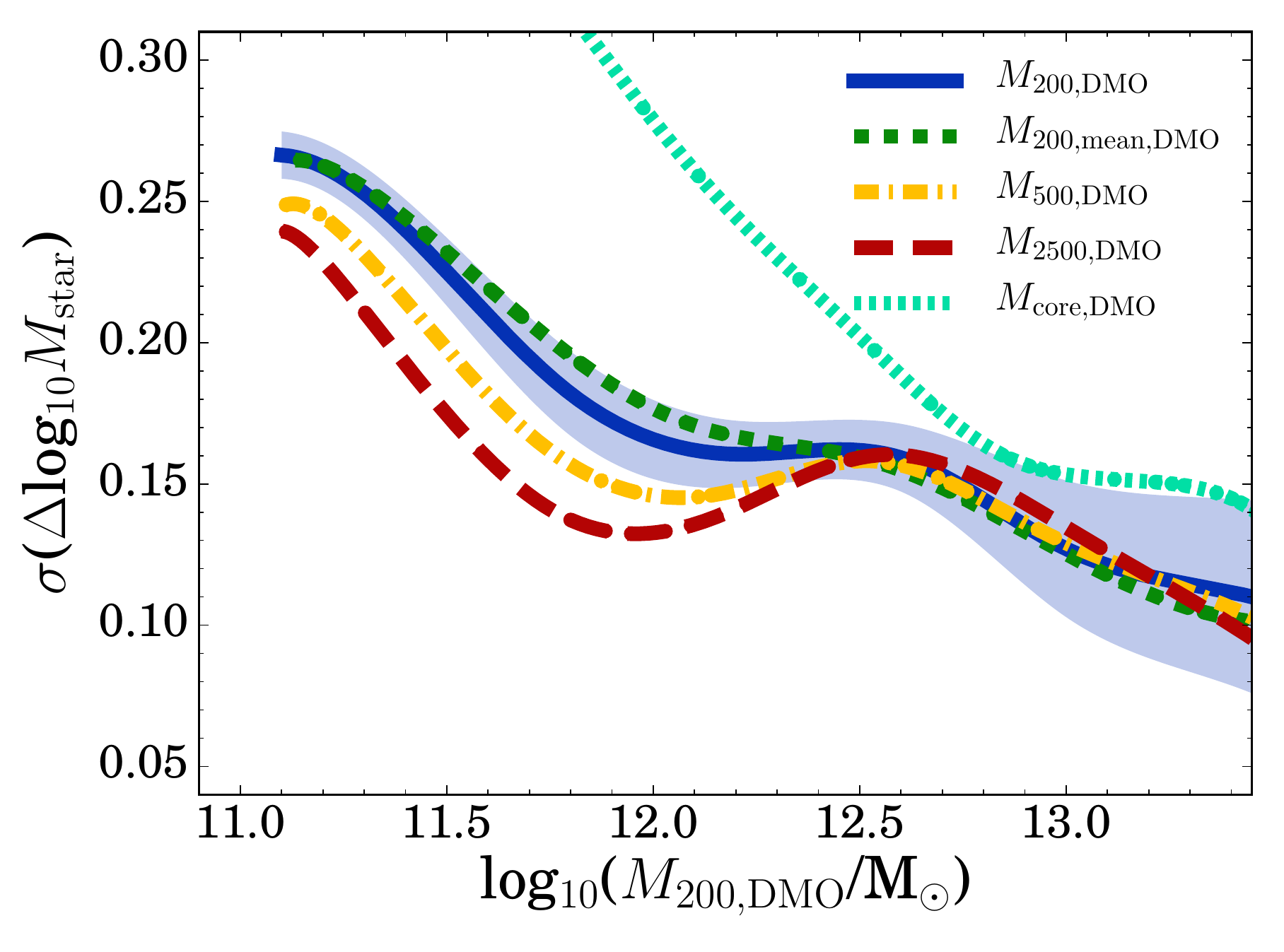}
\caption{\small{As Fig. $\ref{fig:scatter_mstar_haloproperties}$, but now for varying definitions of halo mass. Bins are made in the respective halo property, but we plot the results as a function of the values of $M_{200, \rm DMO}$ corresponding to each bin. The halo mass within $R_{2500, \rm DMO}$ is most strongly related to stellar mass. }}
\label{fig:mhalo_variations}
\end{figure} 

In Fig. $\ref{fig:mhalo_variations}$, we test whether our results depend on our specific choice of halo mass definition. Using $M_{200, \rm mean, DMO}$, which is based on the mass enclosed by the radius within which the mean density is 200 times the mean density of the Universe (as opposed to the critical density used before) results in a slightly larger scatter in the SMHM relation (by $\sim 0.01$ dex, with some dependence on halo mass, see Fig. $\ref{fig:mhalo_variations}$). However, the scatter in the SMHM relation is much larger when using $M_{\rm core, DMO}$.This is surprising, since the core mass is measured at a radius ($r_s$) which is typically half of $R_{2500, \rm DMO}$, and thus more central. A possible explanation is that the NFW fits are  inaccurate in the centres of haloes. Halo mass most accurately predicts stellar mass when it is measured at $R_{500, \rm DMO}$ and $R_{2500, \rm DMO}$, at least for $M_{200, \rm DMO} < 10^{12.5}$ M$_{\odot}$. This is also the case for the binding energy. The halo properties measured at inner radii are more closely related to stellar mass. The same has been shown to hold for galaxy properties other than stellar mass \citep[e.g.][]{Velliscig2014,Zavala2016}.

For comparison, we also have computed the scatter in stellar mass at fixed halo mass when using $M_{200}$ and $V_{\rm max}$ from the baryonic simulation. We find that $\sigma(\Delta{\rm log}_{10} M_{\rm star}$) is $\approx 0.015$ dex smaller at masses $\lesssim 10^{12}$ M$_{\odot}$ when using $M_{200}$ instead of $M_{200, \rm DMO}$. The scatter in stellar mass at fixed rotational velocities is more sensitive to baryonic effects. At masses $\lesssim 10^{12}$ M$_{\odot}$, we find that $\sigma(\Delta{\rm log}_{10} M_{\rm star}$) is $\approx 0.06$ dex smaller when using $V_{\rm max}$ than $V_{\rm max, DMO}$. There are no statistically significant differences between $\sigma(\Delta{\rm log}_{10} M_{\rm star}$) in the baryonic and the DMO simulation at masses $>10^{12}$ M$_{\odot}$.

\begin{figure*}
\centering
\includegraphics[width=17cm]{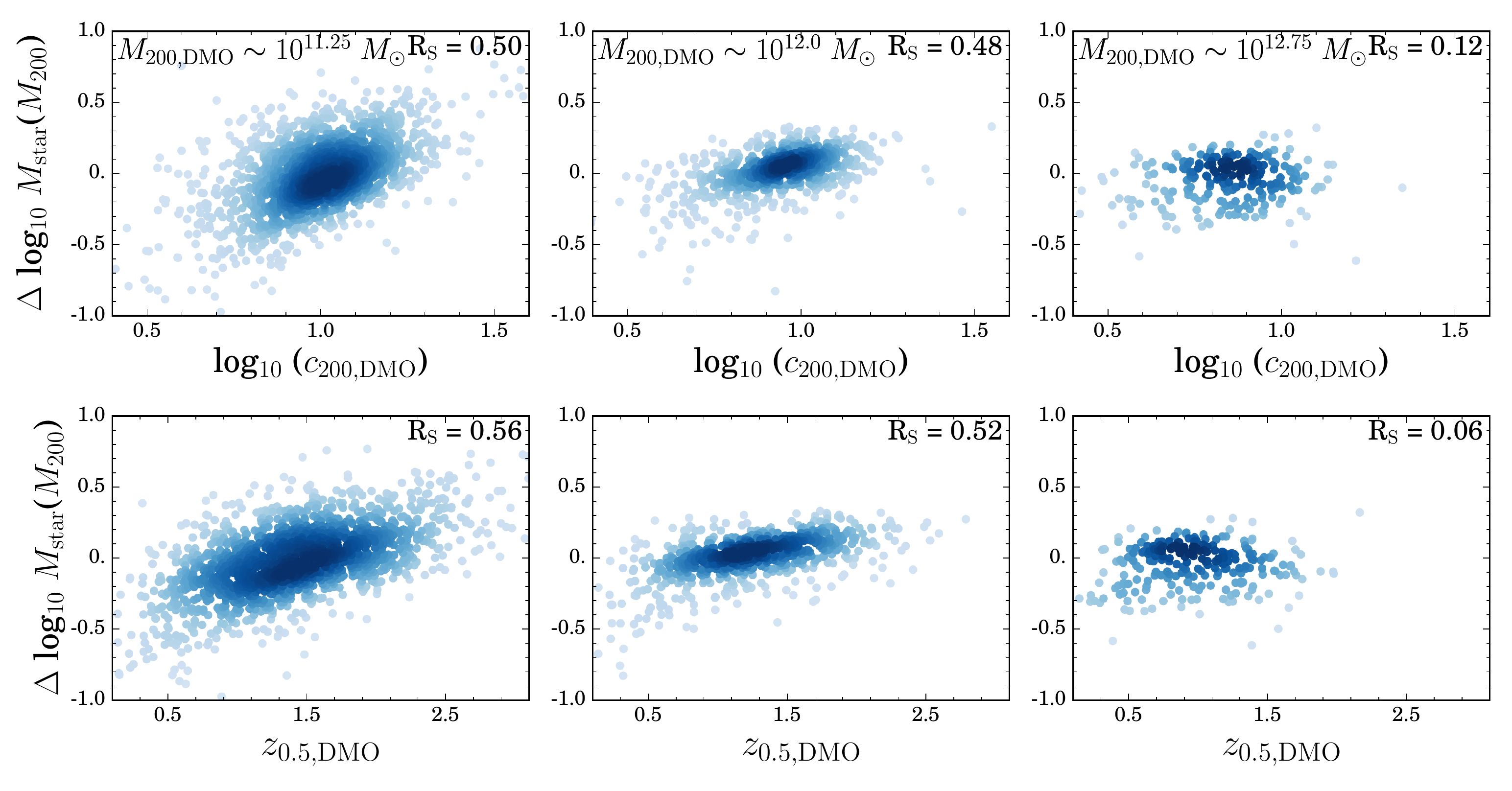}
\caption{\small{{\it Top:} Correlations between the residuals of the stellar mass - halo mass relation ($\Delta$log$_{10} M_{\rm star}$($M_{200, \rm DMO}$)) and DMO halo concentration in the different halo mass ranges (log$_{10}$($M_{200, \rm DMO}/{\rm M_{\odot}}) \approx 11.2, 12.0, 12.6$, from left to right, respectively). The Spearman rank correlation coefficient ($r_s$) is shown in the corner of each panel. A strong correlation can be seen for the low- and intermediate-masses, showing that the scatter in the SMHM relation is partly due to the scatter in halo concentration at fixed mass. {\it Bottom:} Correlations between the residual and the formation time in different halo mass ranges. The results are similar to those for concentration. }}
\label{fig:smhm_res_c200}
\end{figure*} 

\section{Sources of scatter}
In order to understand which processes are the source of scatter in the relations between stellar mass and dark matter halo properties, we investigate the scatter in two scaling relations: $M_{\rm star}- M_{200, \rm DMO}$ and $M_{\rm star}- V_{\rm max}$. We chose halo mass as this is most intuitive and widely used, and $V_{\rm max}$ as this property leads to the most accurate stellar masses (see Fig. $\ref{fig:scatter_mstar_haloproperties}$). In this section we correlate the residuals of these scaling relations with the dimensionless DMO halo properties listed in Table $\ref{tab:properties}$ and discussed in \S 2.3.2. We quantify the strengths of the correlations using the Spearman rank correlation coefficient (R$_s$).

\subsection{Sources of scatter in $M_{\rm star}- M_{200, \rm DMO}$}
We find a strong correlation between the residuals of the SMHM relation and the concentration of the dark matter halo, $c_{200, \rm DMO}$, implying that more concentrated haloes yield higher stellar masses. This effect is strong for both the low- and intermediate-mass ranges (R$_s$ = 0.50, 0.48), see Fig. $\ref{fig:smhm_res_c200}$. We find a weaker correlation for the high halo mass range (R$_s$ = 0.12, P-value 93\%), indicating that there might be different physical processes operating at these halo masses. We have verified that the correlations in the low- and intermediate-mass ranges are not driven by the larger dynamic range in halo concentrations that is sampled thanks to a larger number of objects. By randomly resampling the numbers of galaxies in these mass ranges, such that we get the same number of galaxies as in the high-mass range, we find in all subsamples that R$_s\approx0.5$, with a spread of 0.05. 

The residuals of the relation between $M_{\rm star}$ and both $M_{500, \rm DMO}$ and $M_{2500, \rm DMO}$ are correlated weakly with concentration (not shown). This is because the mass in a more central part of the halo depends on both $M_{200, \rm DMO}$ and concentration. 

We investigate what fraction of the scatter in stellar masses at fixed halo mass is accounted for by concentration. This is done by fitting a linear relation between concentration and the residuals of the SMHM relation, for halo mass bins of 0.4 dex:
\begin{equation}
\Delta {\rm log}_{10} M_{\rm star} (c_{200, \rm DMO}) = a + b \, {\rm log}_{10}(c_{200, \rm DMO}).
\end{equation}
The errors on the normalisation $a$ and slope $b$ of these fits are computed with the jackknife method, as described above. We then fit polynomial relations (with powers up to log$_{10}$($M_{200, \rm DMO}$)$^3$ to the relations in order to obtain the mass dependence of the normalisation and slope, $a$(log$_{10}$($M_{200, \rm DMO}$)), and $b$(log$_{10}$($M_{200, \rm DMO}$)). Then, $\Delta$ log$_{10}$ $M_{\rm star}$($M_{200, \rm DMO}$,$c_{200, \rm DMO}$), the scatter after accounting for concentration, is computed as:
\begin{equation}
\begin{split}
\Delta {\rm log}_{10} M_{\rm star}(M_{200, \rm DMO},c_{200, \rm DMO}) = \Delta {\rm log}_{10} M_{\rm star}(M_{200, \rm DMO})\\
 + a({\rm log}_{10}(M_{200, \rm DMO})) + b({\rm log}_{10}(M_{200, \rm DMO}))\times {\rm log}_{10}(c_{200, \rm DMO}). 
\end{split}
 \end{equation}
 At fixed halo mass, we fold the errors in the normalisation, $\Delta a$, and slope, $\Delta b$, through the errors on the scatter in stellar masses, and obtain the halo mass dependence of the error in the scatter after taking account for concentration, $\sigma$($\Delta$log$_{10} M_{\rm star}$($M_{200, \rm DMO}$,$c_{200, \rm DMO}$)), with a spline interpolation. 

\begin{figure*}
\begin{tabular}{ccc}
\includegraphics[width=5.6cm]{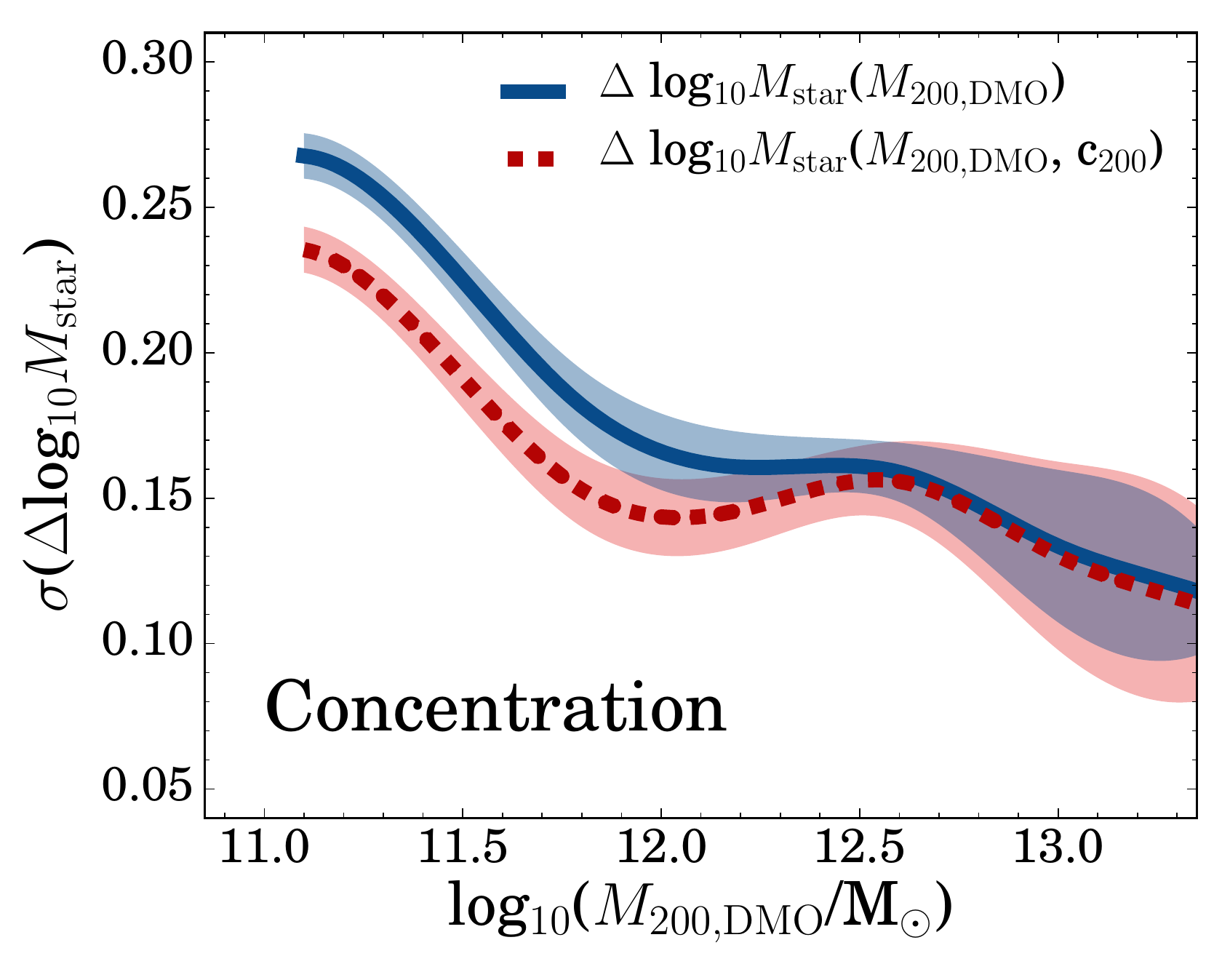}&
\includegraphics[width=5.6cm]{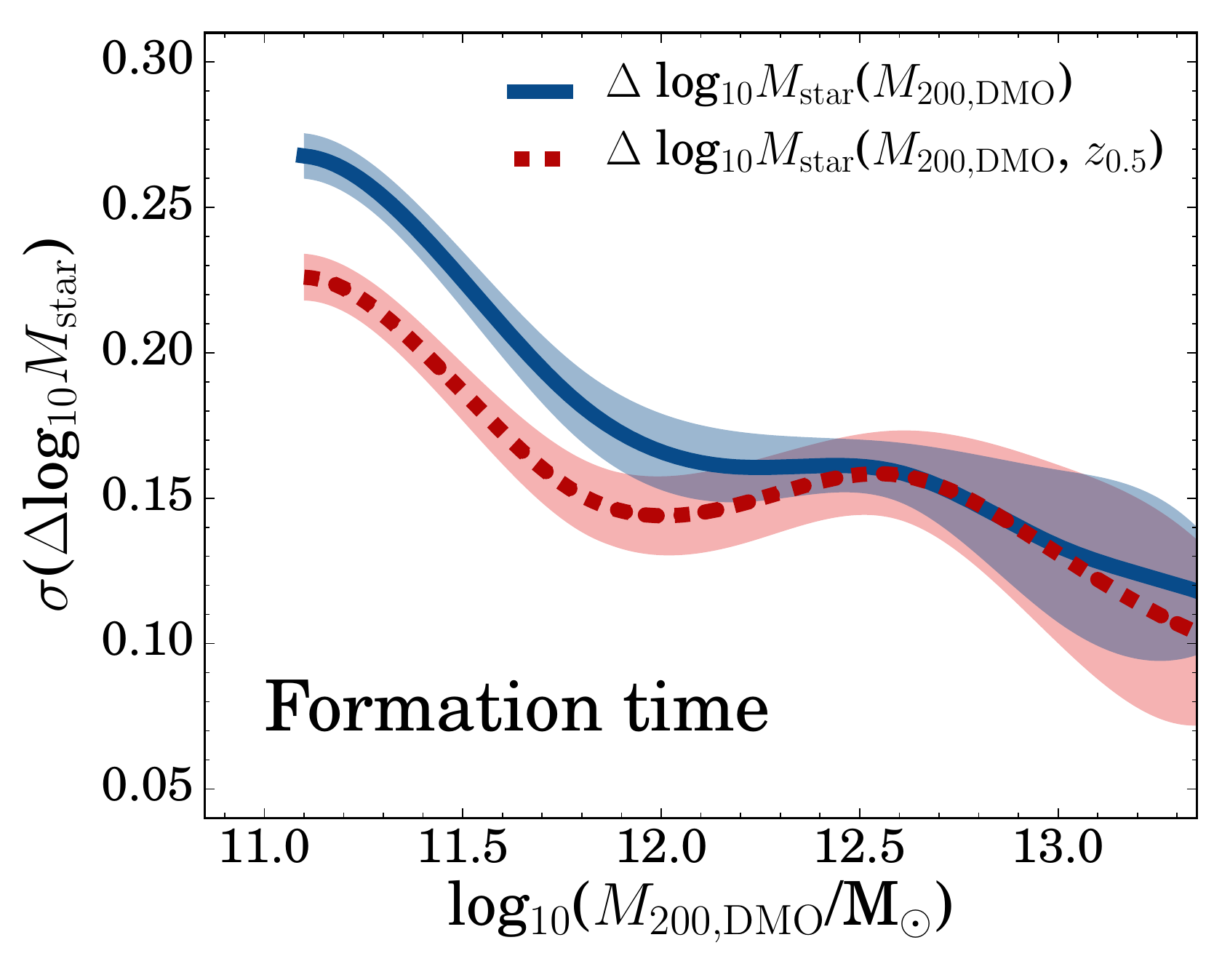}&
\includegraphics[width=5.6cm]{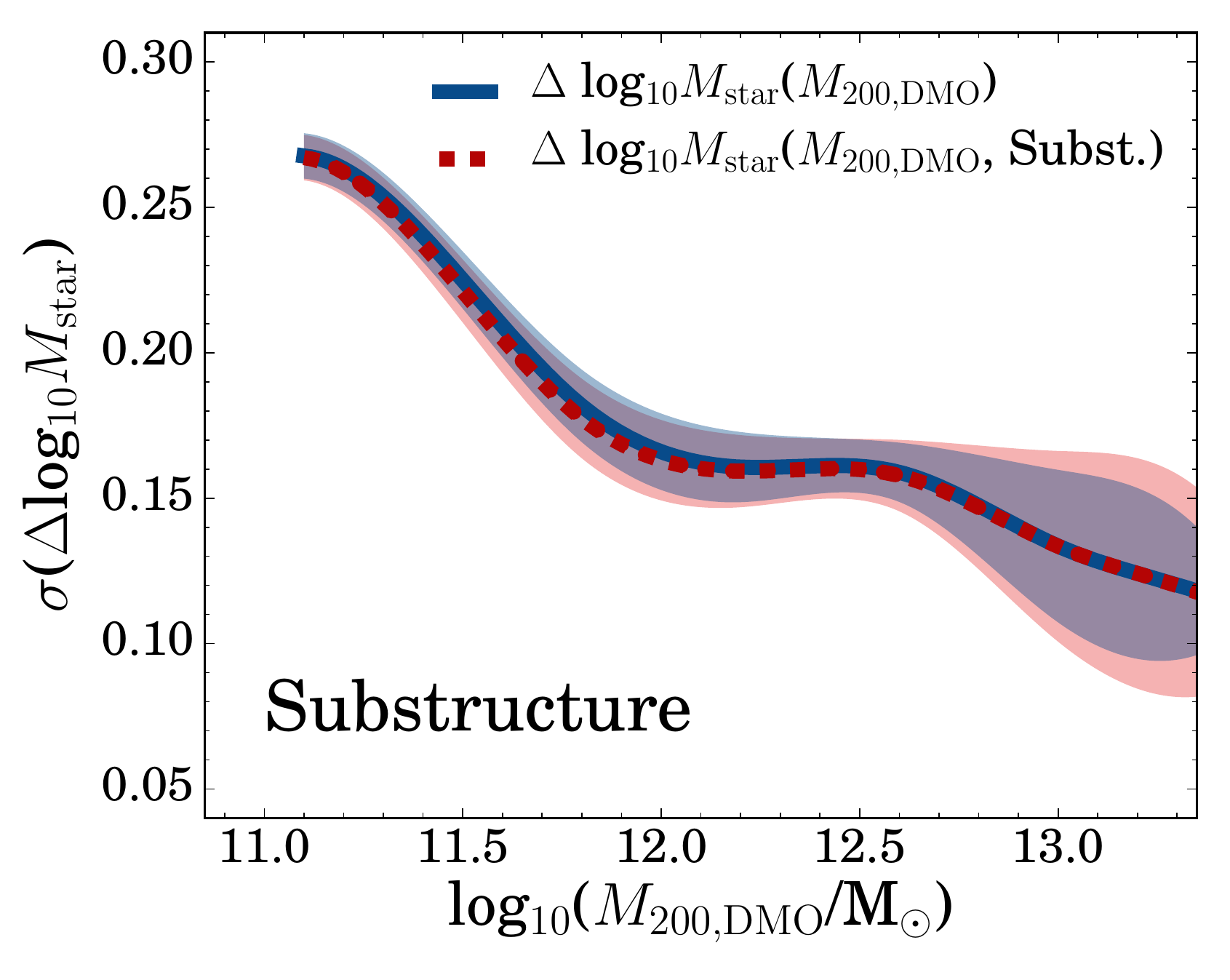}\\
\includegraphics[width=5.6cm]{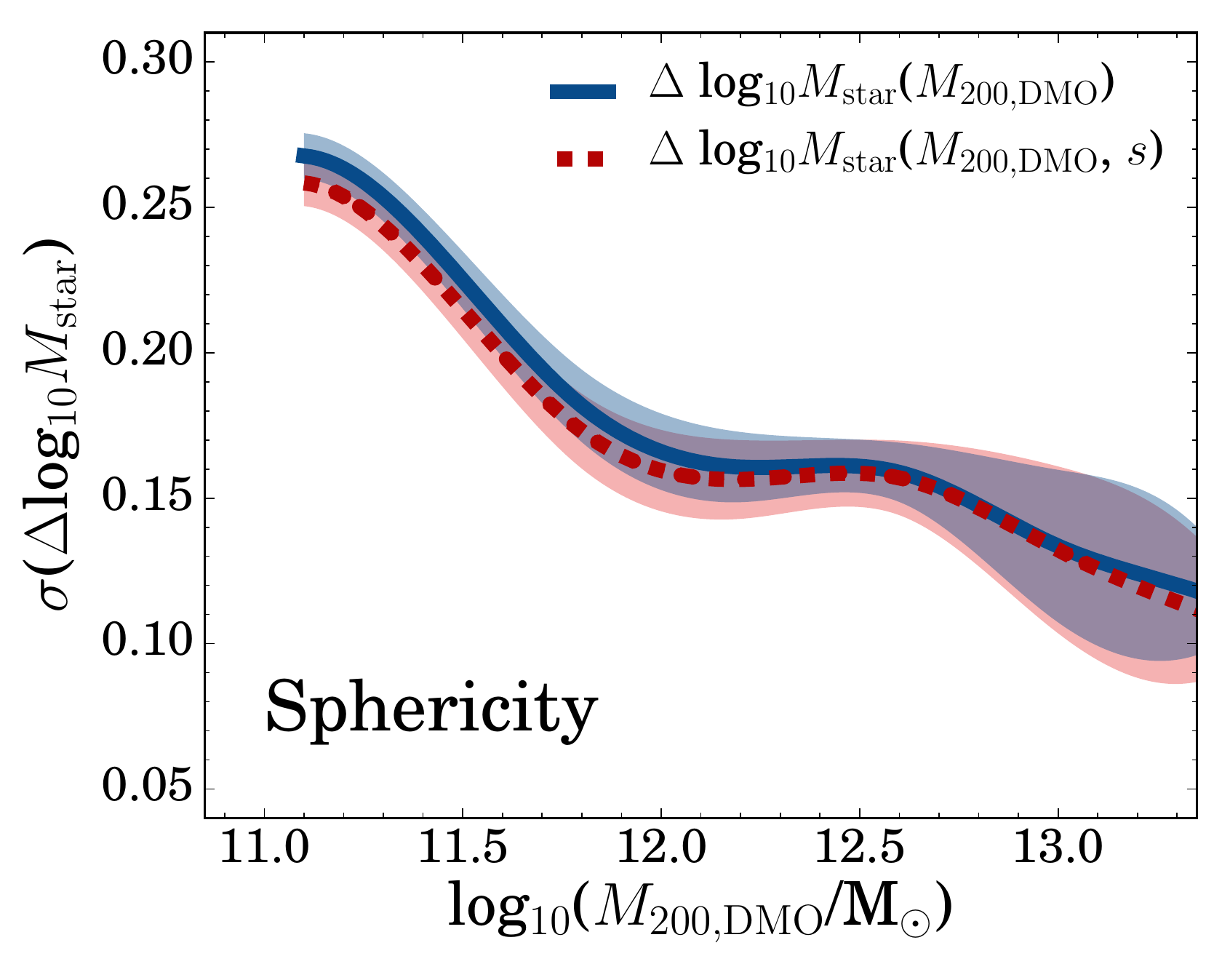}&
\includegraphics[width=5.6cm]{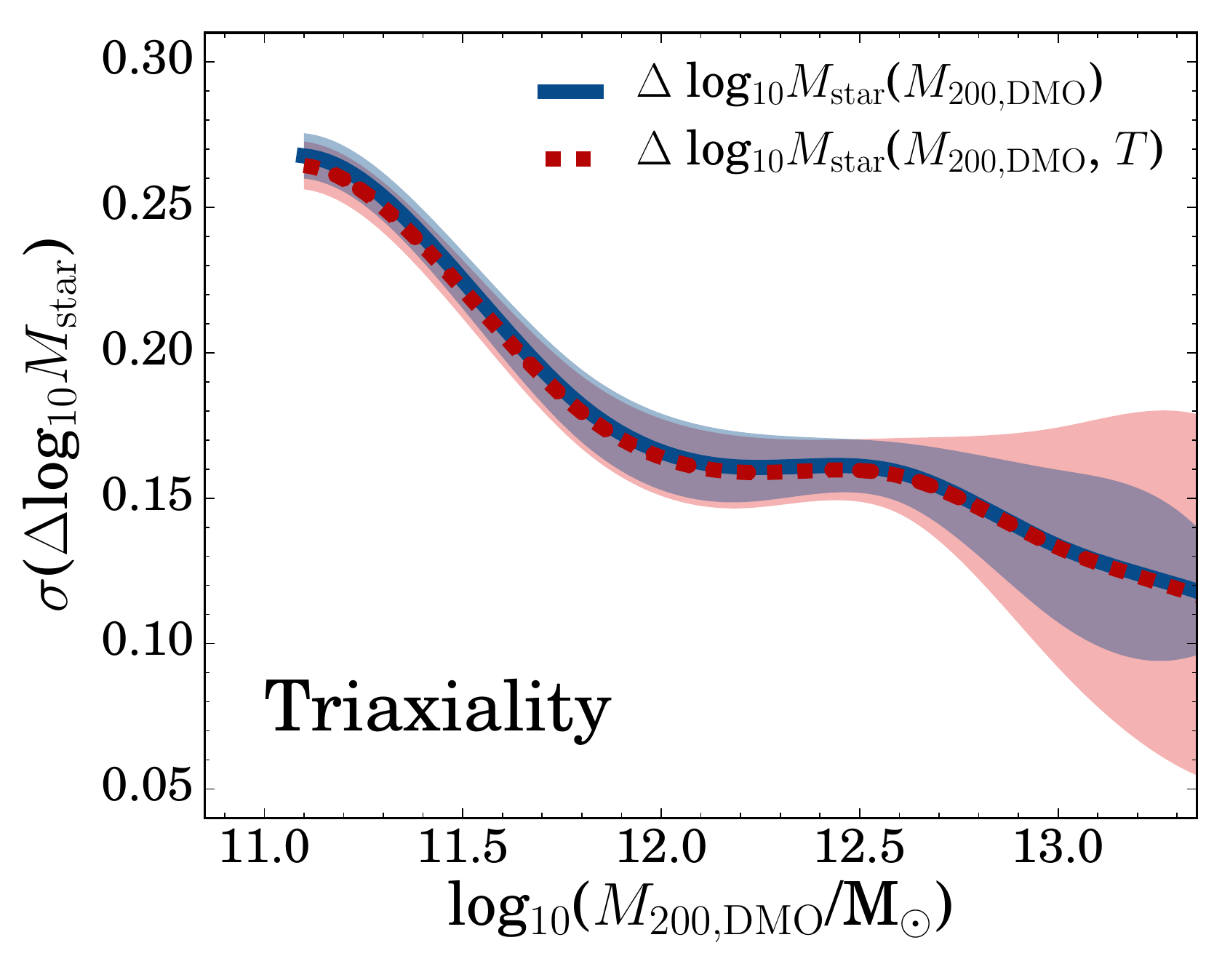}&
\includegraphics[width=5.6cm]{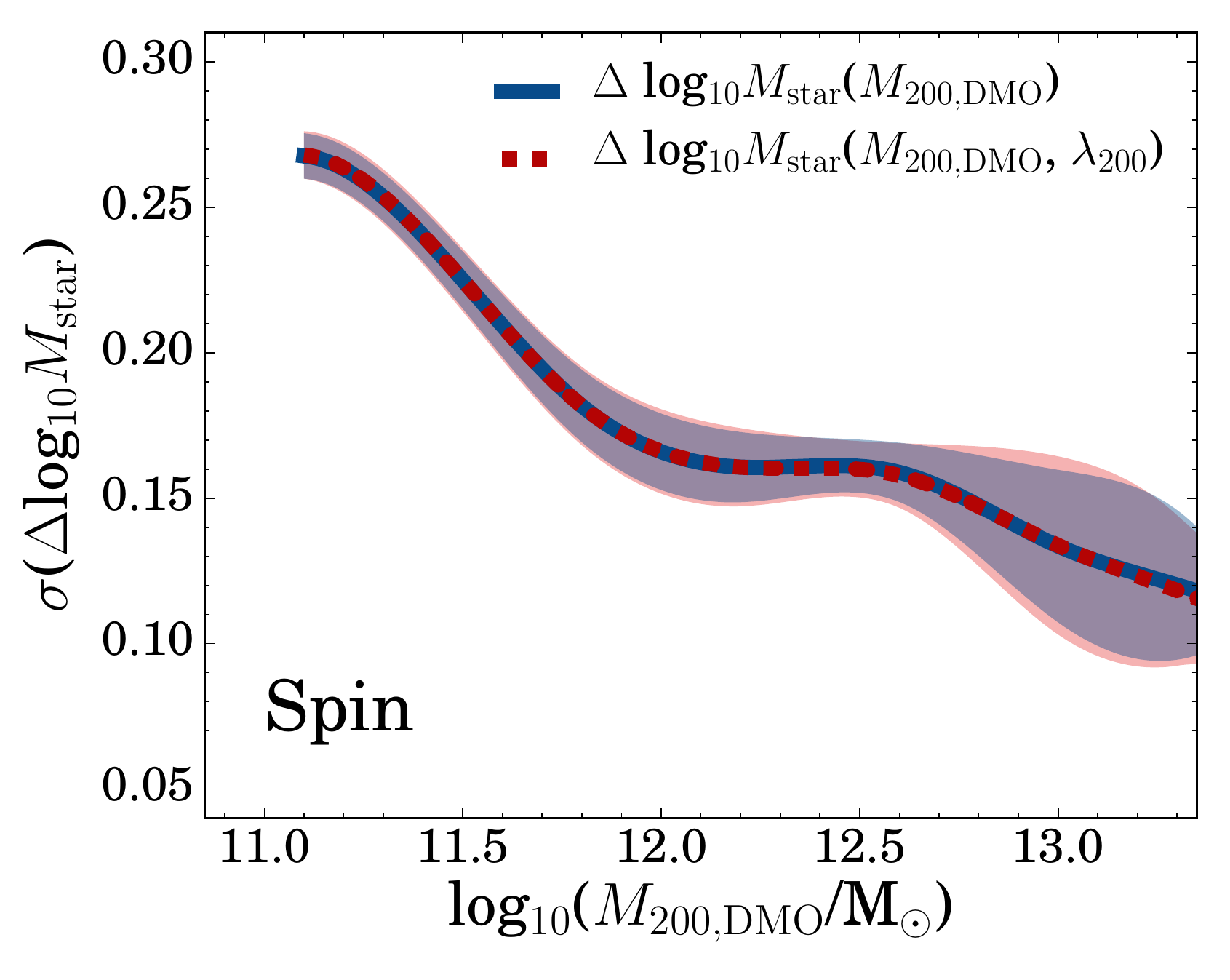}\\
\includegraphics[width=5.6cm]{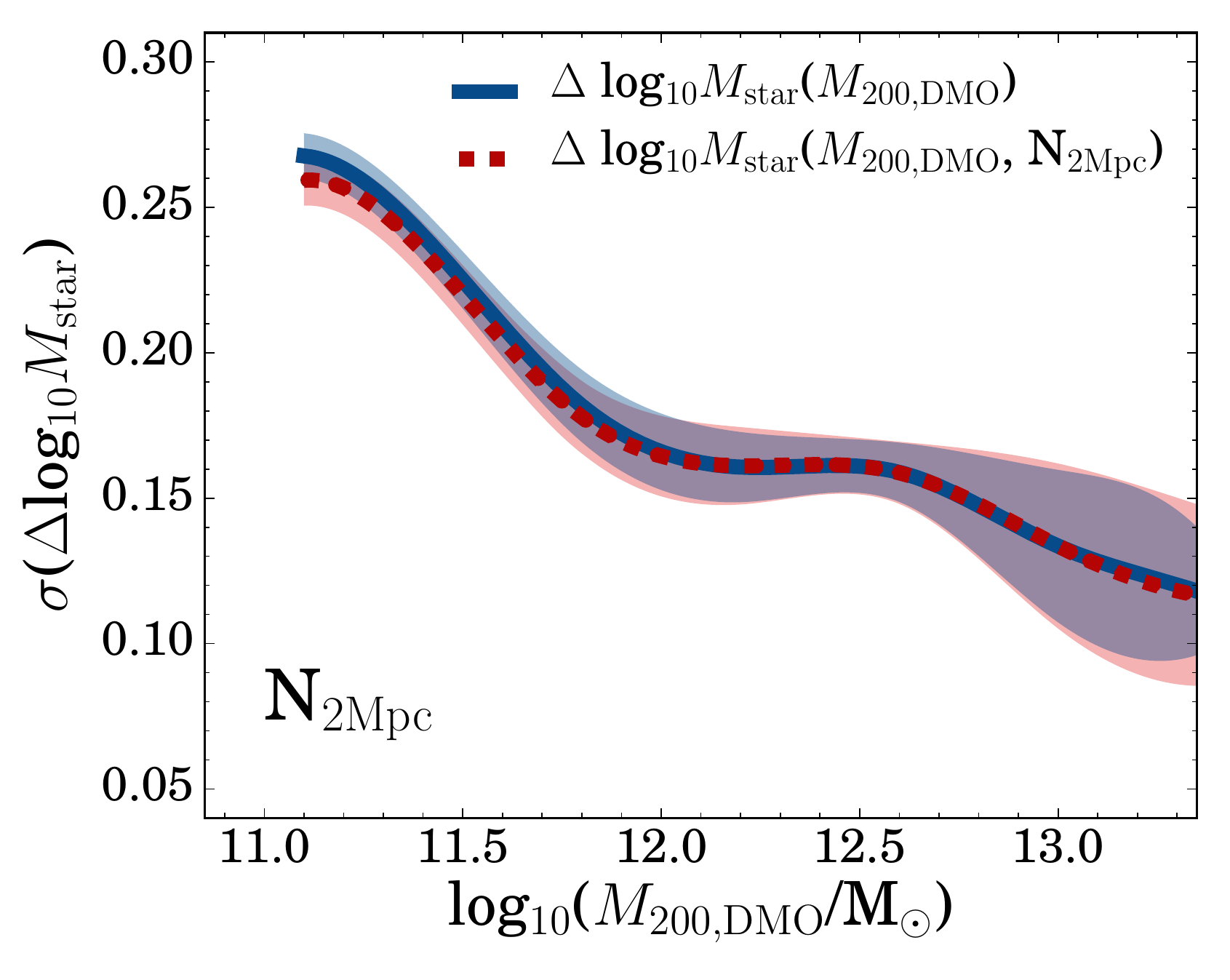}&
\includegraphics[width=5.6cm]{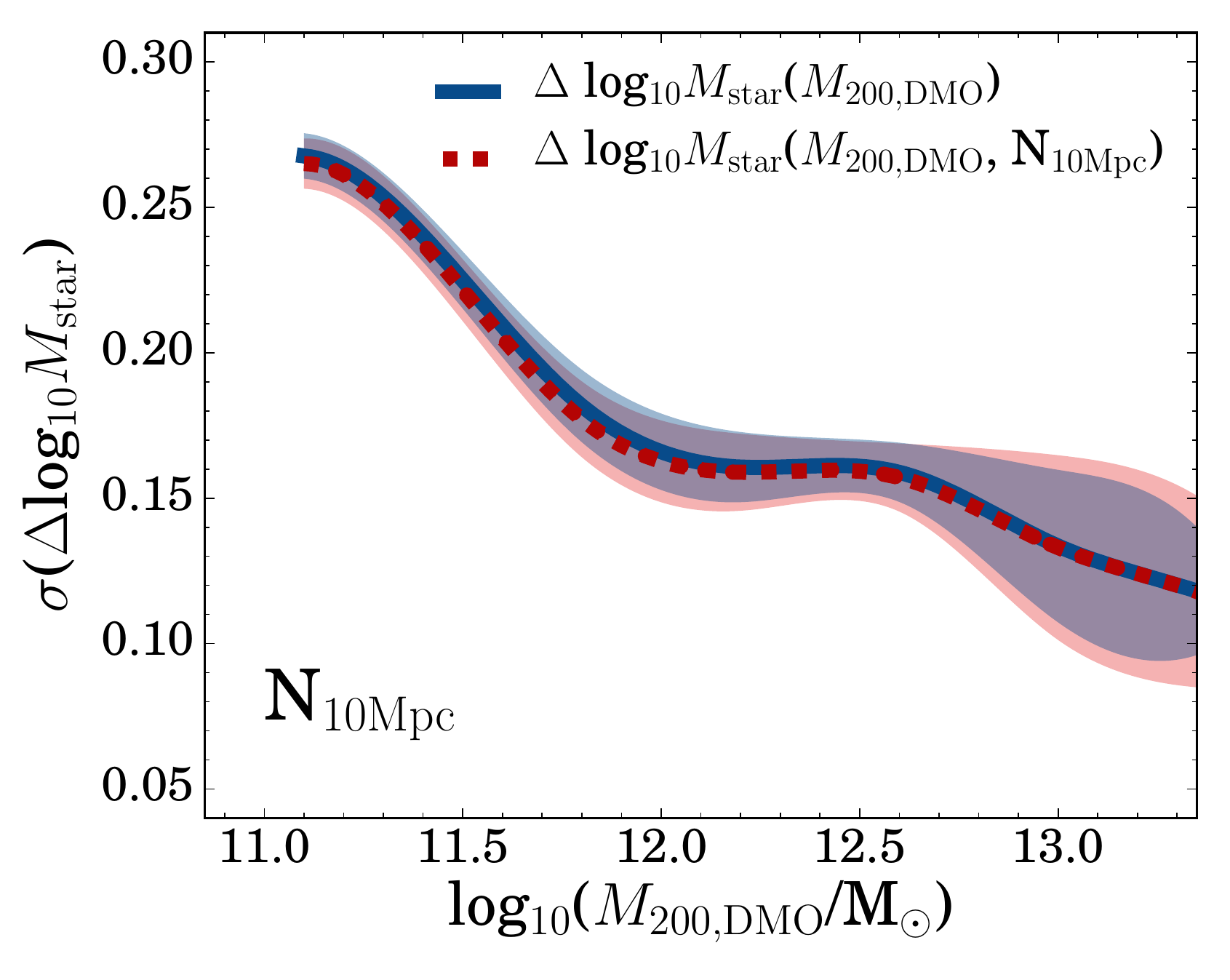}&

\end{tabular}
\caption{\small{Scatter in the difference between the true and predicted stellar mass as a function of DMO halo mass, before and after using a dimensionless DMO halo property in addition to mass, in blue and red, respectively. Each panel corresponds to a different property. The shaded regions indicate the 1$\sigma$ uncertainty. Only $c_{200, \rm DMO}$ and $z_{0.5, \rm DMO}$ are responsible for a statistical improvement in the scatter in stellar masses.}}
\label{fig:range_spearman}
\end{figure*}

The result is shown in the top-left panel of Fig. $\ref{fig:range_spearman}$. At the lowest halo masses, 0.03 dex of the scatter in stellar masses is accounted for by concentration, while this is lower at higher masses. For $M_{200, \rm DMO} > 10^{12.5}$ M$_{\odot}$ the inclusion of concentration does not reduce the scatter in stellar mass, again indicating that different physical processes are at play \citep[i.e.][]{Tinker2016b}.

It is interesting to note that \cite{Jeeson-Daniel2011} found from a principal component analysis of DMO simulations that halo concentration is the most fundamental halo property, being strongly related to many other dimensionless halo properties, and that halo mass only sets the scale of a system. This is consistent with our results, as we find that once the scale of the halo is factored out (by studying residuals at fixed halo mass), concentration is correlated with stellar mass. Furthermore, \cite{BoothSchaye2010} find that the black hole masses in their hydrodynamical simulation are set by halo mass with a secondary dependence on concentration, similar to our results for stellar mass, leading them to conclude that the halo binding energy is the most fundamental halo property in setting black hole masses. It could be that halo binding energy also determines stellar masses, as it is for example more difficult to drive galactic winds out of a galaxy in a halo with a steeper potential well. At the highest masses the correlation with binding energy may weaken because star formation is quenched and galaxies grow predominantly through mergers.

\begin{figure}
\centering
\includegraphics[width=8.6cm]{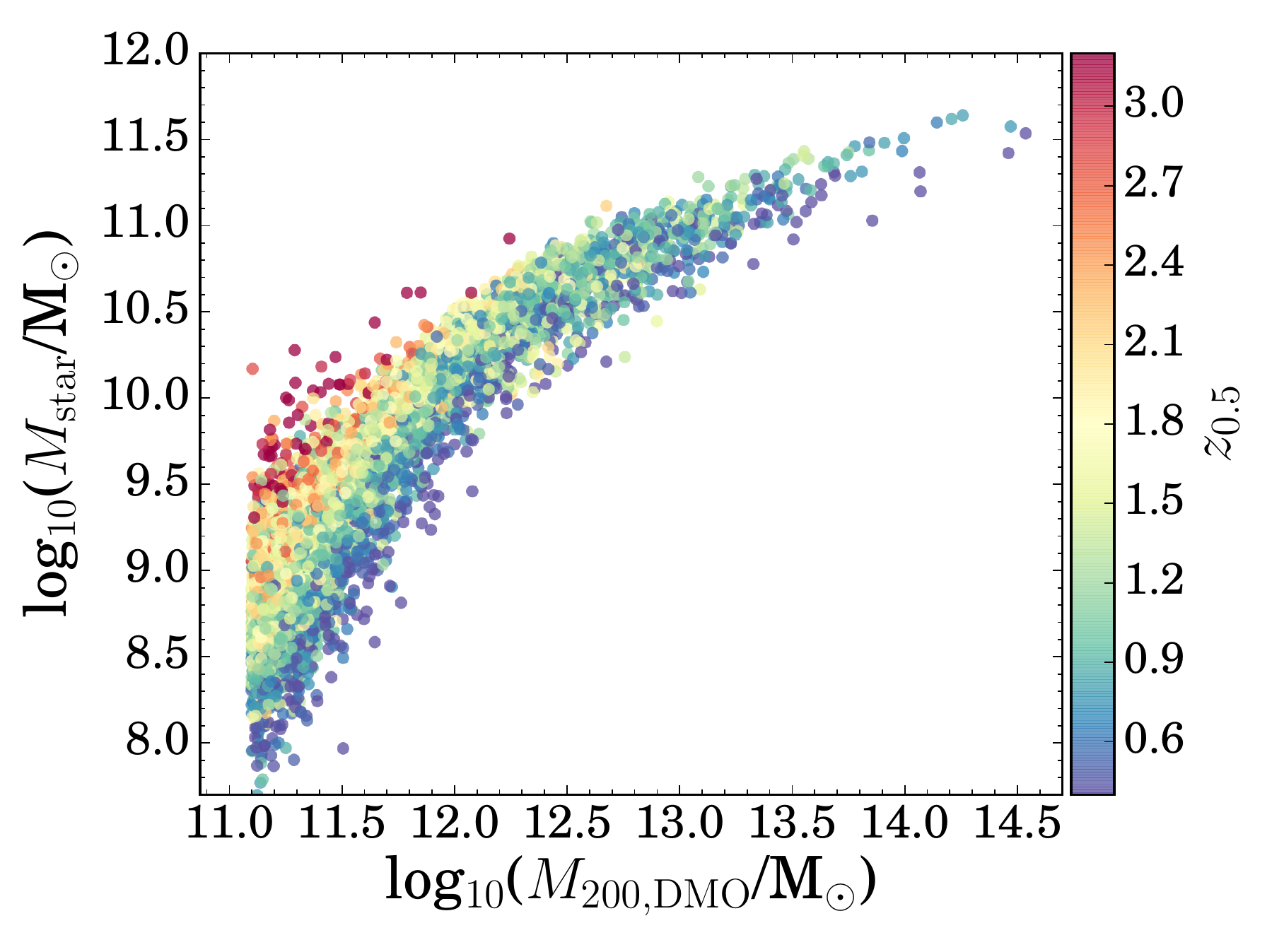}
\caption{\small{Stellar mass - halo mass relation in the EAGLE simulation colour-coded with the formation redshift of the halo. At low and intermediate halo masses, an earlier formation time corresponds to a higher stellar mass at fixed halo mass. This correlation is not seen at the highest masses. The transition occurs at masses slightly above $10^{12}$ M$_{\odot}$, where the SMHM relation also flattens.}}
\label{fig:SMHM_colorz05}
\end{figure} 

Since concentration is strongly correlated with formation time \citep[e.g.][]{Wechsler2002,Zhao2009,Jeeson-Daniel2011,Ludlow2014,Correa2014}, we expect galaxies with a large stellar mass at fixed halo mass to have formed earlier. We indeed find that the residuals of the SMHM relation correlate with $z_{0.5,\rm DMO}$, particularly for halo masses below $\sim10^{12}$ M$_{\odot}$, as illustrated in Fig. $\ref{fig:smhm_res_c200}$. In Fig. $\ref{fig:range_spearman}$ it can be seen that $z_{0.5,\rm DMO}$ is responsible for roughly the same amount of scatter in stellar masses, as concentration is. This is further illustrated in Fig. $\ref{fig:SMHM_colorz05}$, which shows that haloes that form galaxies relatively efficiently generally form earlier. 

Hence, another explanation for the correlation between concentration (and formation time) and the residuals of the SMHM relation is that haloes with a higher concentration started forming stars earlier and will thus be able to reach a higher stellar mass by a fixed redshift.

For halo masses $>10^{12}$ M$_{\odot}$ there is almost no correlation between formation time and the residuals of the SMHM relation. As was the case for concentration, a possible explanation for this is that in the most massive haloes stars have formed earlier than the assembly of their final halo, which is generally known as down-sizing \citep[e.g.][]{Cowie1996,DeLucia2006}.

Fig. $\ref{fig:range_spearman}$ shows that no halo property considered here, other than concentration and formation time, is responsible for the scatter in stellar mass at fixed halo mass. We find that there are weak correlations (R$_s \approx 0.3$) between the residuals of the SMHM relation and sphericity, substructure and N$_{2 \rm Mpc, DMO}$ for masses $<10^{12}$ M$_{\odot}$. However, these might be explained by correlations between these quantities and concentration \citep[e.g.][]{Jeeson-Daniel2011}. Since accounting for the concentration (or formation time) reduces the scatter in stellar mass by only $\lesssim 0.04$ dex, most of the scatter in the SMHM relation cannot be explained in terms of variations in the DMO halo properties. 

It is interesting to measure how strongly the residuals of the SMHM relation are correlated with the concentration of the dark matter halo as measured in the full baryonic simulation. This correlation is much stronger for all halo mass ranges (R$_s$ = 0.77, 0.79 and 0.47) than the correlation between the DMO concentration and the residuals of the SMHM relation (R$_s$ = 0.50, 0.48 and 0.12). This implies that a higher concentration is both a cause of and an effect from efficient galaxy formation. For a given halo mass, efficient cooling (and thus star formation) leads to a higher concentration \citep[e.g.][]{Blumenthal1986,Duffy2010,Schaller2014profile}. However, the concentration from the dark matter only version of the simulation can only be a cause of more efficient galaxy formation. Thus, for a given halo mass, a higher dark matter halo concentration will lead to a higher stellar mass, which then results in an even more concentrated dark matter halo in the full baryonic simulation.

\subsubsection{Robustness of results and varying definitions of concentration and formation time} 
The fact that halo concentration is itself weakly correlated with halo mass \citep[e.g.][]{Navarro1997,AvilaReese1999,Duffy2008}, with the parametric form $c_{200, \rm DMO} \propto M_{200, \rm DMO}^B$, with $B\approx -0.1$, can influence our results. We remove this dependence by correlating the residuals of the SMHM relation with the residuals of the $c_{200, \rm DMO}$-$M_{200, \rm DMO}$ relation obtained with the non-parametric method. We find that this does not change the Spearman coefficient for the correlation between $\sigma(\Delta{\rm log}_{10} M_{\rm star}$) and concentration by more than 0.02, regardless of halo mass range.

We varied our definition of the concentration, as it might be important how we define the viral radius and because the use of an NFW profile to obtain the concentration might bias the results. Definitions that were tested are based on the circular velocity in the dark matter only version at various radii: $V_{\rm max}$/$V_{200, \rm DMO}$, $V_{\rm max}$/$V_{500, \rm DMO}$ and $V_{\rm max}$/$V_{2500, \rm DMO}$. However, all correlate slightly less or equally strong with the residuals of the SMHM relation than is the case for $c_{200, \rm DMO}$. This suggests that our definition of concentration is close to optimal.
A similar result is found when we vary the definition of formation time. The correlations between formation time and the residuals of the SMHM relation are slightly weaker if other assembly mass-fractions than 0.5 are chosen (we tested fractions of 0.33, 0.66 and 0.75). This indicates that our somewhat arbitrary choice of a mass fraction of 0.5 is close to optimal. 

\begin{figure*}
\begin{tabular}{cccc}
\includegraphics[width=4.11cm]{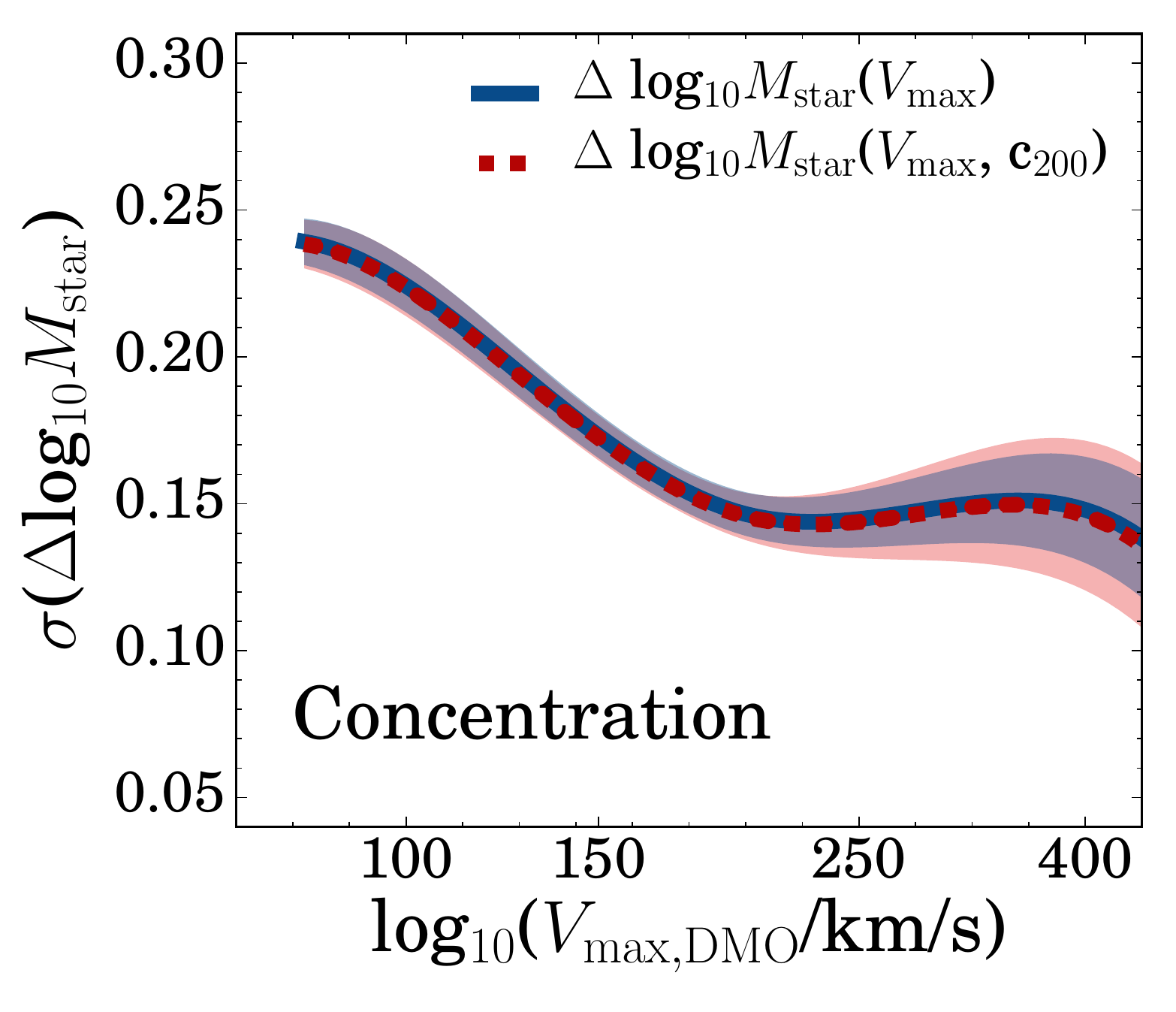}&
\includegraphics[width=4.11cm]{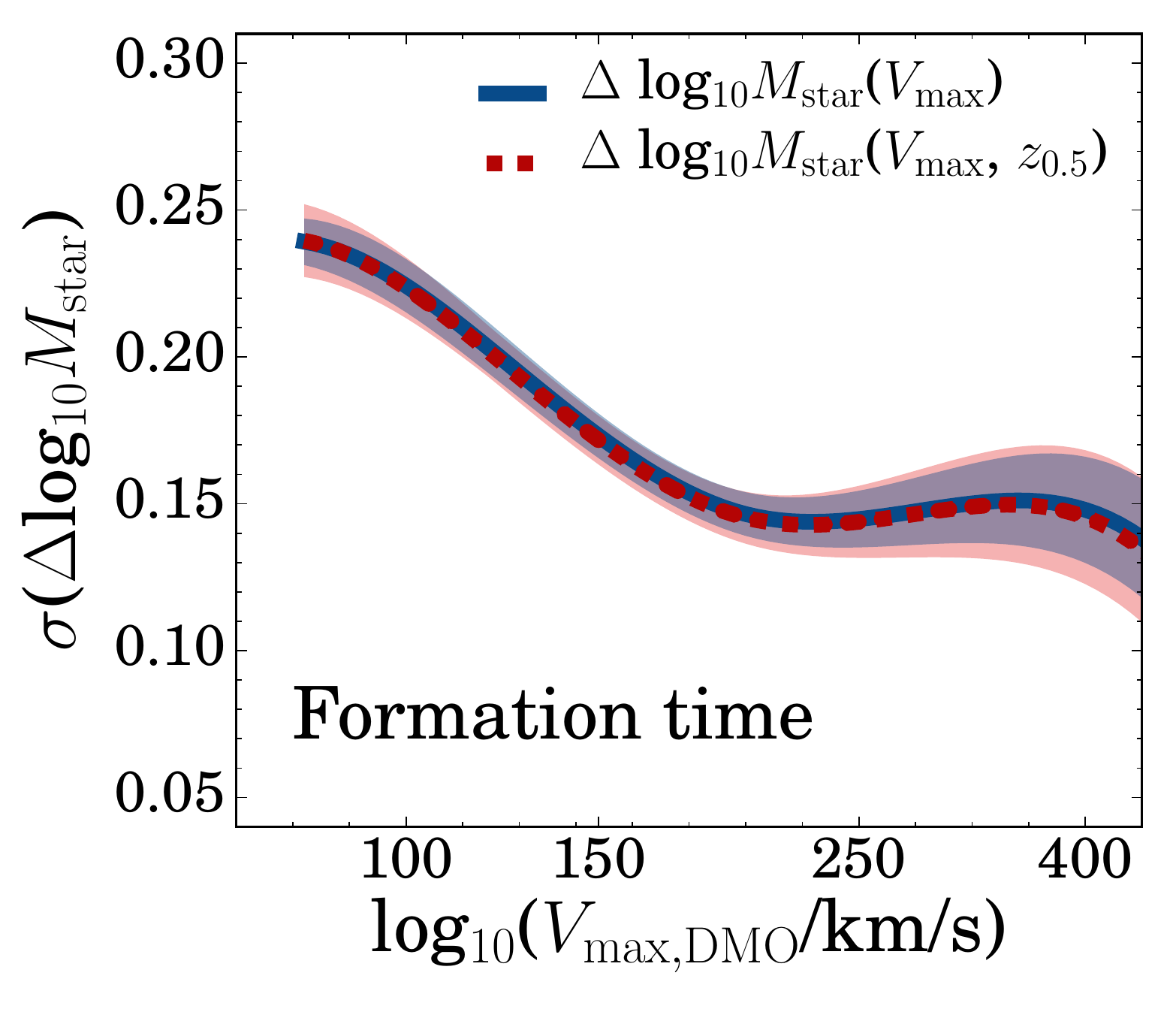}&
\includegraphics[width=4.11cm]{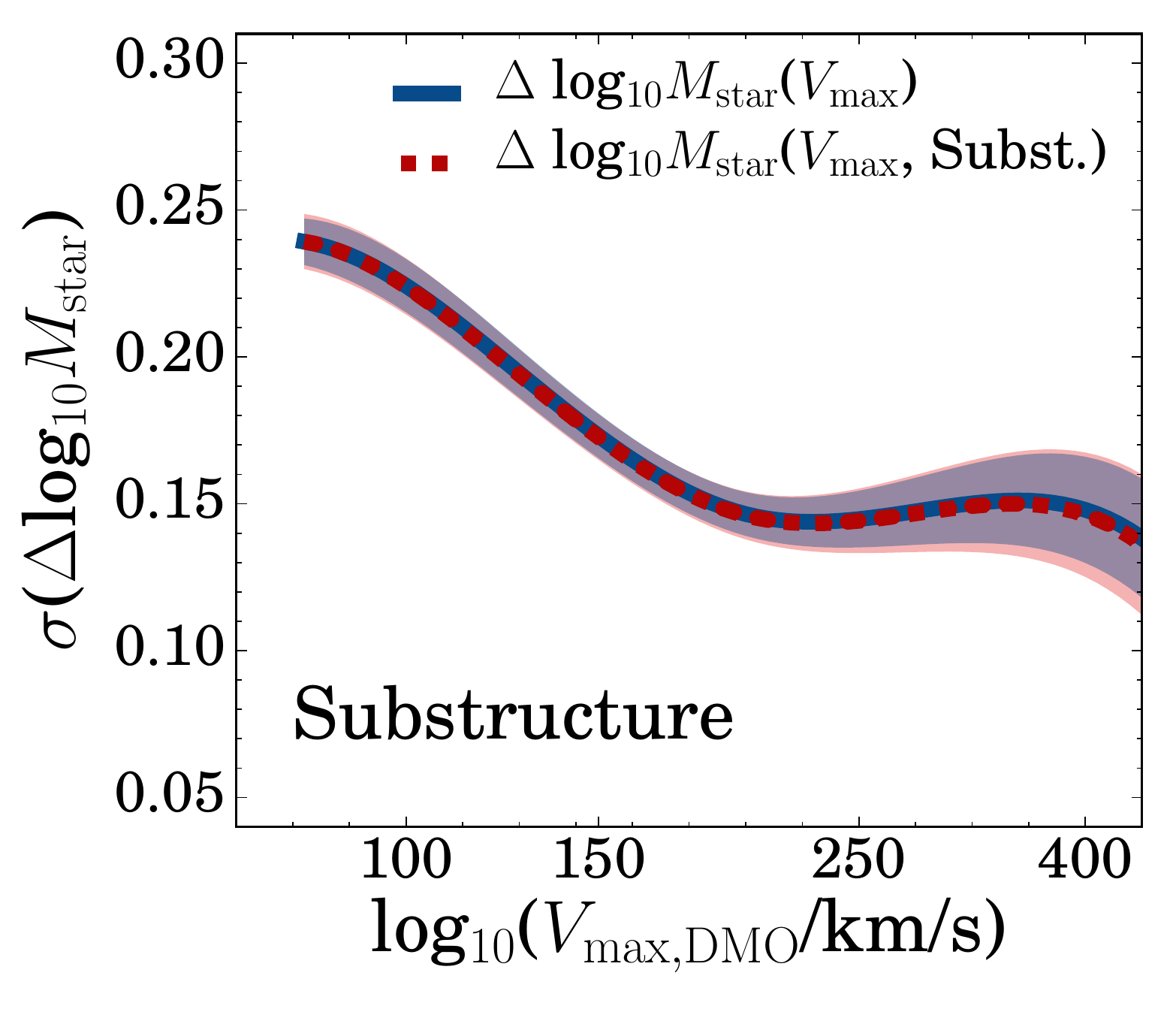}&
\includegraphics[width=4.11cm]{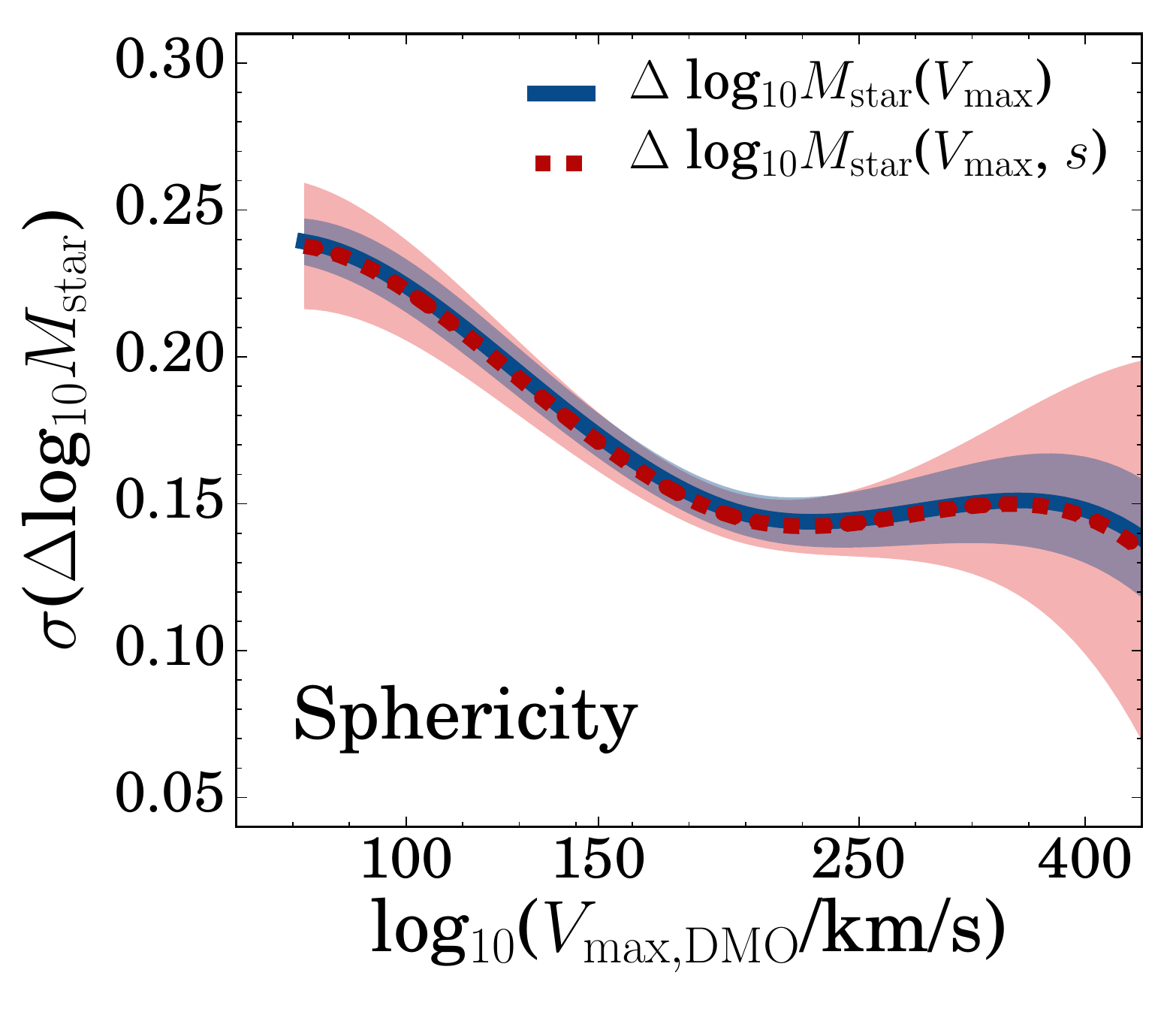}\\
\includegraphics[width=4.11cm]{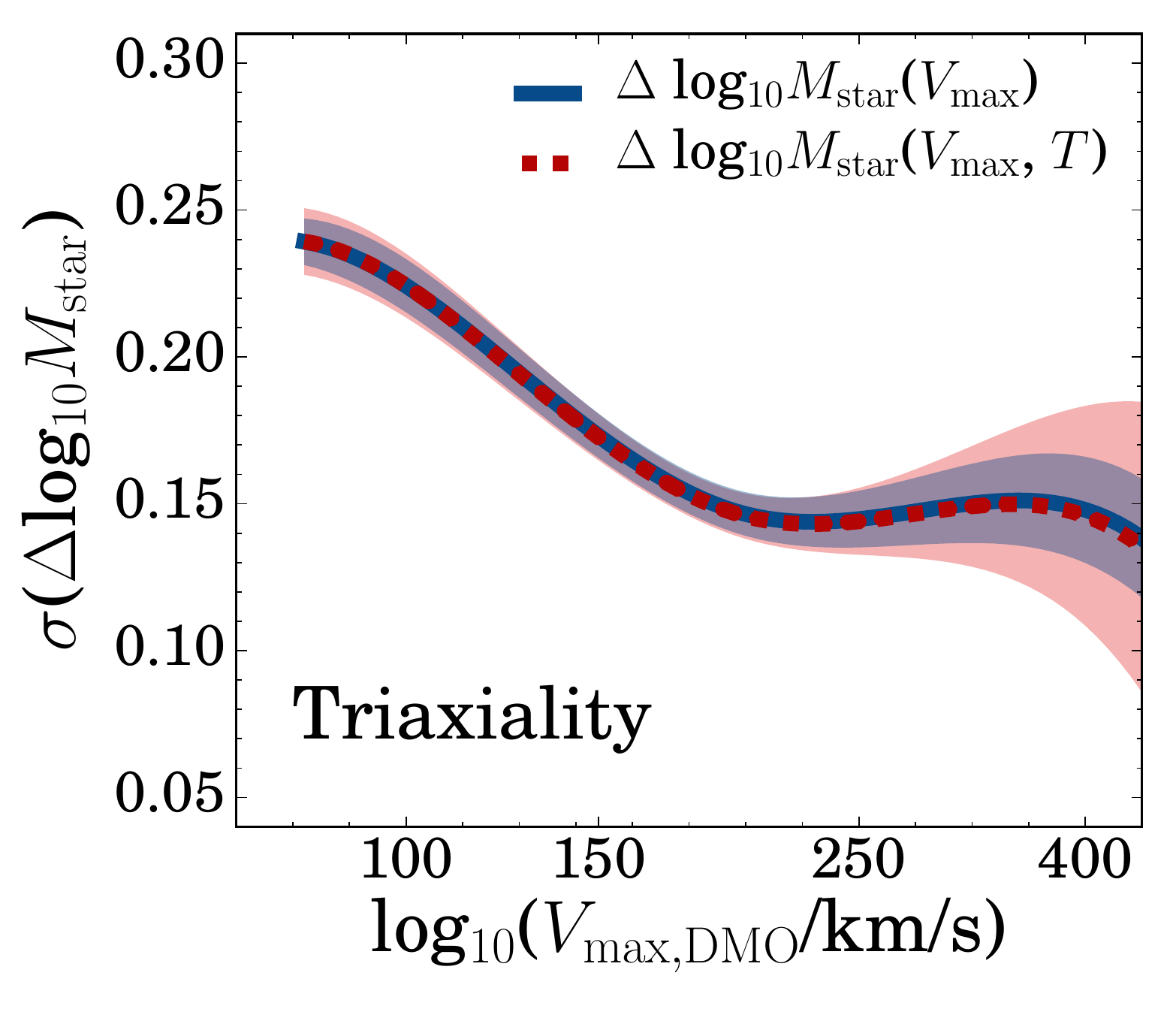}&
\includegraphics[width=4.11cm]{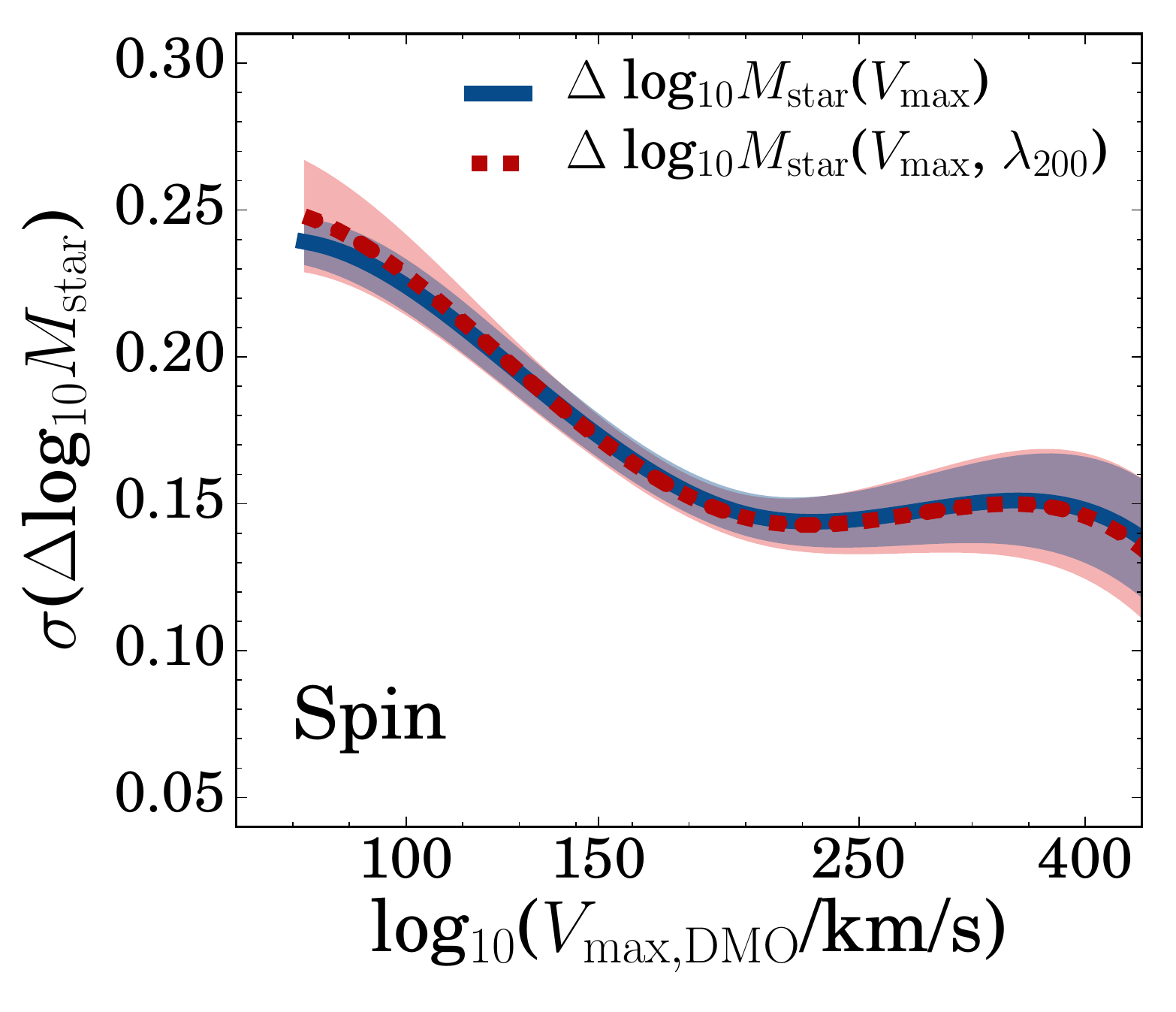}&
\includegraphics[width=4.11cm]{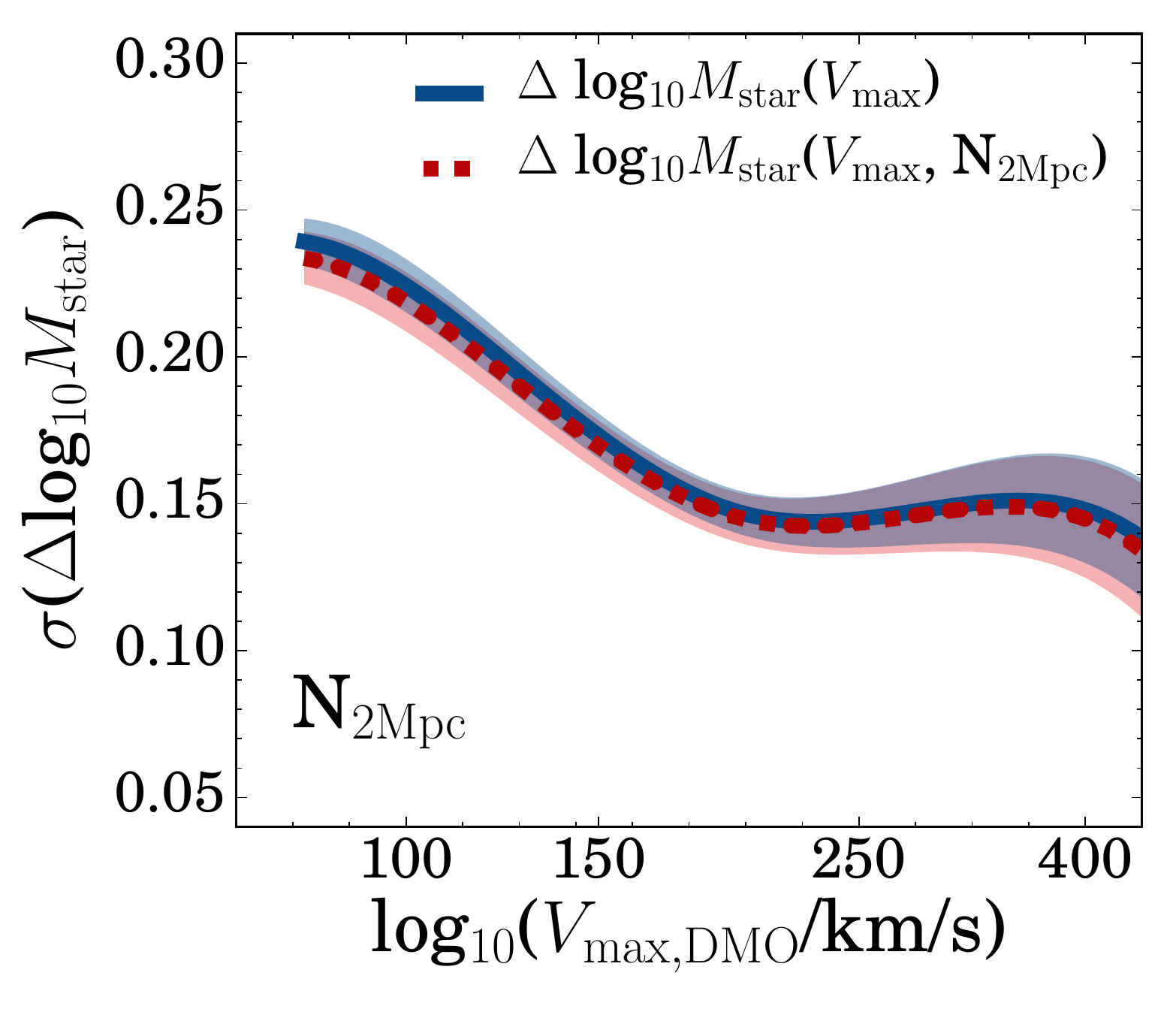}&
\includegraphics[width=4.11cm]{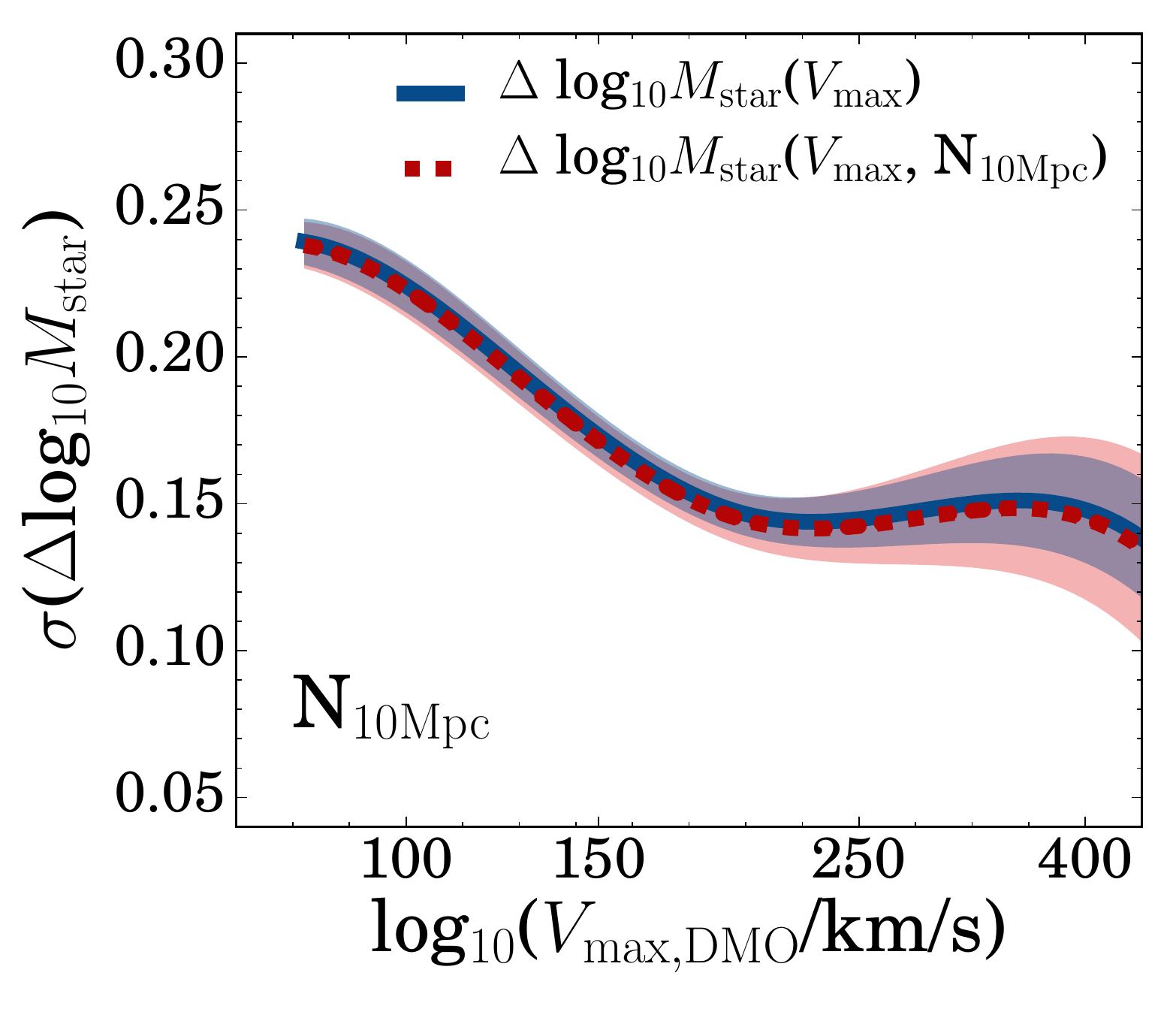}\\

\end{tabular}
\caption{\small{As Fig. $\ref{fig:range_spearman}$ but now with the scatter in stellar mass as a function of $V_{\rm max, DMO}$ instead of $M_{200, \rm DMO}$. The addition of a dimensionless halo property to $V_{\rm max, DMO}$ does not result in statistically more accurate stellar masses.}}
\label{fig:range_spearman_vmax}
\end{figure*} 

We also test the effect of selecting only relaxed haloes, using the definition from \cite{Duffy2008}. This means that we only select haloes for which the distance between the centre of mass and the most bound particle is smaller than 0.07 times the virial radius. The fractions of relaxed halos in the low, intermediate and high halo mass range are 0.65, 0.55 and 0.52, respectively. For the highest halo mass range, we find that there are no differences. For the low- and intermediate-masses, the correlation between the scatter and concentration becomes slightly weaker (R$_s = 0.45$, $0.35$, respectively). This is expected since the spread in concentration will be smaller, as concentration is correlated with relaxedness \citep[e.g.][]{Jeeson-Daniel2011}.

To test the impact of recent interactions between haloes, we remove central galaxies which have been satellite galaxies in the recent past ($<3$ Gyr) or will become satellites between $z=0.1$ and $z=0.0$ (note that we carry out our analysis at $z=0.1$). While some of these galaxies have either some of the highest or lowest stellar masses for their halo mass, there is little difference statistically. The $\sigma(\Delta{\rm log}_{10} M_{\rm star}$) decreases by $\lesssim0.01$ dex for all mass ranges and the correlation between $\sigma(\Delta{\rm log}_{10} M_{\rm star}$) and formation time becomes slightly stronger for the low and intermediate mass ranges (R$_s$ = 0.57 and 0.55, respectively), and similarly for concentration.

\subsection{Sources of scatter in $M_{\rm star}- V_{\rm max, DMO}$}
In \S 3 we showed that $V_{\rm max, DMO}$ is somewhat more closely related to the stellar mass of central galaxies than $M_{200, \rm DMO}$ is. However, there is still significant scatter in the $M_{\rm star}- V_{\rm max, DMO}$ relation. Therefore, we now investigate whether any dimensionless halo properties correlate with the residuals of the relation between stellar mass and $V_{\rm max, DMO}$. Note that $V_{\rm max, DMO}$ is very closely related to $E_{2500,\rm DMO}$, with a scatter of only 0.05 dex. 

Similarly as for $M_{200, \rm DMO}$ in Fig. $\ref{fig:range_spearman}$, Fig. $\ref{fig:range_spearman_vmax}$ shows the 1$\sigma$ spread in the residuals of the $M_{\rm star}- V_{\rm max, DMO}$ relation as a function of $V_{\rm max, DMO}$, before and after correcting for the dependence on a dimensionless halo property. None of the investigated halo properties reduce the scatter in stellar mass. Indeed, we find no strong (|R$_s|>0.3$) correlations between the residuals of the $M_{\rm star}$ - $V_{\rm max, DMO}$ relation and dimensionless DMO halo properties. The fact that we find no correlation with concentration or formation time, means that the additional scatter in the SMHM relation due to concentration is already accounted for by $V_{\rm max}$, which is related to both halo mass and concentration. This is also shown in Fig. $\ref{fig:mstarreconstruct}$, which compares the spread in stellar mass as a function of halo mass, where the stellar mass is computed either from $M_{200, \rm DMO}$, $V_{\rm max, DMO}$ or $E_{2500,\rm DMO}$ alone, or from $M_{200, \rm DMO}$ and either $c_{200, \rm DMO}$ or $z_{0.5\rm, DMO}$. Note that while we have binned in $V_{\rm max, DMO}$ and $E_{2500,\rm DMO}$, we show the halo masses corresponding to those bins respectively. By comparing the green curve (for $V_{\rm max}$) with the dashed curves (using $M_{200, \rm DMO}$ and an additional property), it is clear that $M_{200, \rm DMO}$ performs less well than the other predictors.

\section{A Parametric description for predicting stellar masses}
As described in \S 4.1, up to 0.04 dex of scatter in stellar masses at fixed halo mass is attributed to variations in formation times and concentrations (where we measured the scatter in the SMHM relation with the non-parametric method). In this section, we use the parametric method to obtain fitting functions for stellar mass as a function of halo mass and concentration or formation time.

\subsection{Halo mass and formation time}
We correct the stellar mass at fixed $M_{200, \rm DMO}$ using a fit between the scatter in the SMHM  ($\Delta M_{\rm star}$($M_{200, \rm DMO}$)) and DMO formation time. As before, we use a simple linear least squares fit between the residuals of the SMHM and $z_{0.5, \rm DMO}$, which results in:

\begin{equation}
\label{eq3}
\begin{split}
\Delta {\rm log}_{10} M_{\rm star}(M_{200, \rm DMO},z_{0.5,\rm DMO})= \\ a({\rm log}_{10} M_{200, \rm DMO}/{\rm M}_{\odot})\, z_{0.5,\rm DMO} 
+ b({\rm log}_{10} M_{200, \rm DMO}/{\rm M}_{\odot}).
\end{split}
\end{equation}

\noindent When including all galaxies (such that we average over all halo masses), we find best fitting parameters $a=0.22^{+0.01}_{-0.01}$ and $b= -0.31^{+ 0.01}_{-0.01}$. 

However, we have seen that the dependence on formation time varies with halo mass. We therefore need to fit the parameters $a$ and $b$ in a mass-dependent way. This mass dependence is obtained in the same way as we obtain the mass dependence of the scatter in the SMHM relation, which was described in \S 2.5.

The relations between the slope and normalisation of Eq. $\ref{eq3}$ and halo masses are fit with a cubic relation.
\begin{equation}
\label{eq4}
a(X) = -196.005 + 49.262\, X - 4.107 \, X^2 + 0.114 \, X^3,
\end{equation}
where we define $X={\rm log}_{10} (M_{200, \rm DMO}/{\rm M}_{\odot})$, and
\begin{equation}
\label{eq5}
 b(X) = 154.322 - 39.571 \,X + 3.357 \, X^2 - 0.094 \, X^3, 
\end{equation}

Combining Equations $\ref{eq2}$, $\ref{eq3}$, $\ref{eq4}$ and $\ref{eq5}$, we find that we can predict stellar masses at $z=0.1$ to a precision of $\approx0.12-0.22$ dex from DMO halo properties with the following equation:
\begin{equation}
\label{eq6}
{\rm log}_{10} (M_{\rm star}/{\rm M}_{\odot}) = \alpha-e^{\beta \,X +\gamma}+ a(X) \, z_{0.5,\rm DMO} + b(X),
\end{equation}
where $\alpha$, $\beta$ and $\gamma$ are listed in the first line of Table $\ref{tab:fits}$. We note that the errors on the fits for $a(X)$ and $b(X)$ are large at $M_{200, \rm DMO} > 10^{12.5}$ M$_{\odot}$. Above that halo mass, $a(X)$ and $b(X)$ should therefore be set to zero. 

Using Eq. $\ref{eq6}$ instead of Eq. $\ref{eq2}$ reduces the 1$\sigma$ scatter in the difference between predicted stellar masses and true stellar masses from 0.26 to 0.23 dex and from 0.16 to 0.14 dex in the low-mass and intermediate-mass ranges respectively by construction ($a=b=0$ for $M_{200, \rm DMO} > 10^{12.5}$ M$_{\odot}$. This is illustrated in Fig. $\ref{fig:mstarreconstruct}$, where we compile the scatter in the difference between true and predicted stellar mass as a function of DMO halo mass for various parametric fits.

\subsection{Halo mass and concentration}
Although formation time correlates slightly better with the residuals of the SMHM relation than concentration does, we can also use concentration as a secondary parameter to obtain more accurate stellar masses. We repeat the same steps as the previous section by using ${\rm log}_{10}(c_{200, \rm DMO})$ in stead of $z_{0.5,\rm DMO}$. The benefit of using ${\rm log}_{10}(c_{200, \rm DMO})$ is that we do not rely on the merger tree, and therefore only require the simulation output of a DMO simulation at a single snapshot. For the simulation output at $z=0.1$, we obtain the following equation:
\begin{equation}
\label{eq7}
d(X) = -399.944 + 100.358\, X - 8.341 \, X^2 + 0.230 \, X^3,
\end{equation}
and,
\begin{equation}
\label{eq8}
e(X) = 296.274 -75.165\, X + 6.307 \, X^2 - 0.175 \, X^3.
\end{equation}
Finally, this results in:
\begin{equation}
\label{eq9}
{\rm log}_{10} (M_{\rm star}/{\rm M}_{\odot}) = \alpha-e^{\beta \,X +\gamma}+ d(X) \, {\rm log}_{10}c_{200, \rm DMO} + e(X),
\end{equation}
where the relevant $\alpha$, $\beta$ and $\gamma$ are listed in the first line of Table $\ref{tab:fits}$. We note again that above $M_{200, \rm DMO} > 10^{12.5}$ M$_{\odot}$, $d(X)$ and $e(X)$ are set to zero because of the large errors. 

\begin{figure}
\includegraphics[width=8.6cm]{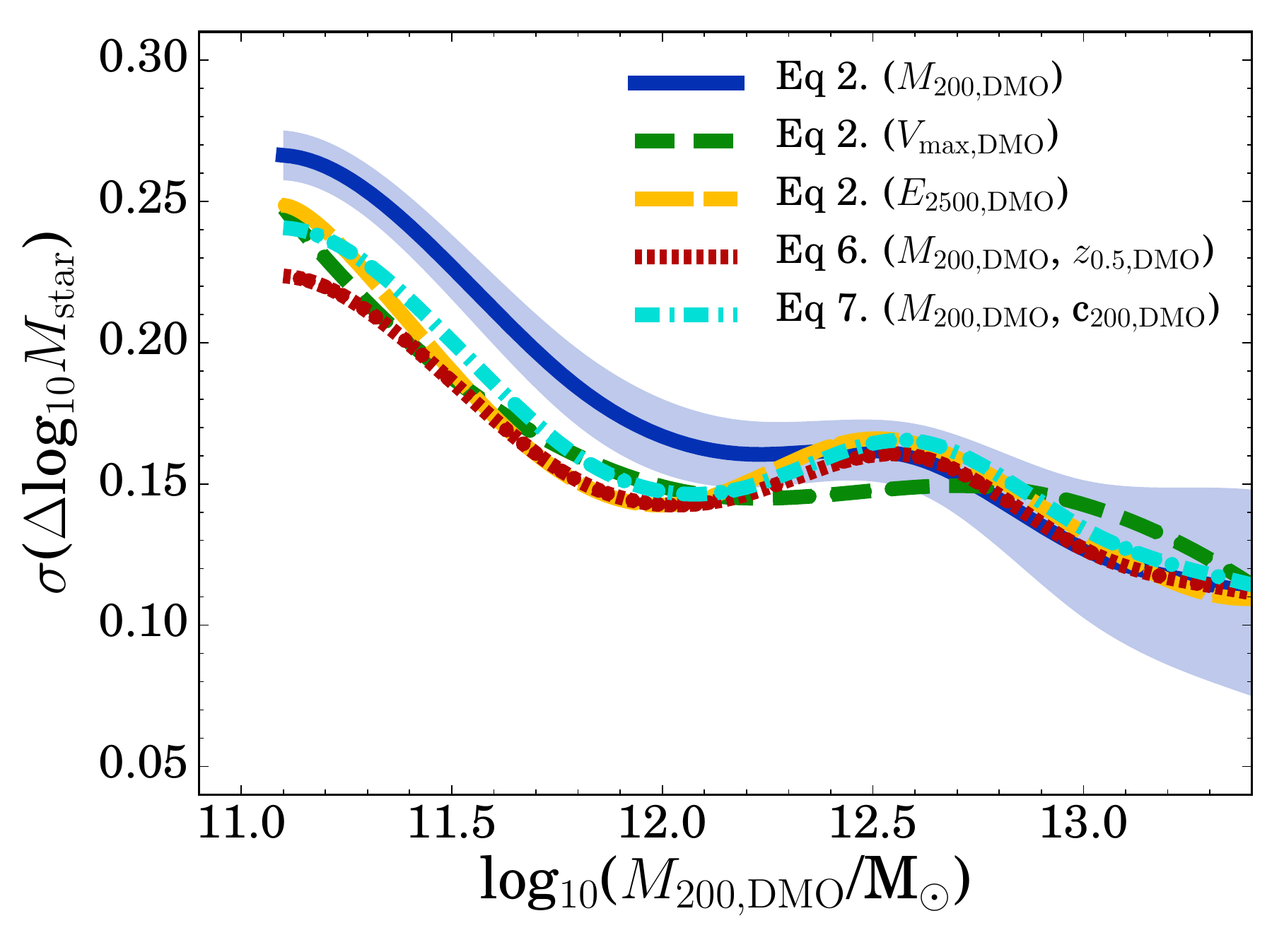} 
\caption{\small{Scatter in the difference between true and predicted stellar mass from various parametric fits as a function of $M_{200, \rm DMO}$. To first order, the stellar mass can be computed using halo masses and Eq. $\ref{eq2}$. A second order correction based on the relation between the scatter in the SMHM relation and formation time or concentration is applied using either Eq. $\ref{eq6}$ or Eq. $\ref{eq7}$. Since the scatter in the SMHM relation does not correlate with formation time at the highest halo masses, the scatter is only reduced for halo masses below $10^{12.6}$ M$_{\odot}$. It can be seen that using formation time is slightly more robust than using concentration. The scatter is then very similar to the scatter in stellar mass as a function of $V_{\rm max}$. }}
\label{fig:mstarreconstruct}
\end{figure}

When comparing the statistical corrections to stellar masses using formation time or concentration in Fig. $\ref{fig:mstarreconstruct}$, it is clear that using the formation time is only marginally better. One possible reason that the formation time performs slightly better than $c_{200, \rm DMO}$ at low halo masses could be that there is some other scatter in $c_{200, \rm DMO}$ at low halo mass that is due to numerical noise because the number of dark matter particles available to constrain the fitted NFW-profile is small.

\begin{figure*}
\centering
\begin{tabular}{cc}

\includegraphics[width=8.6cm]{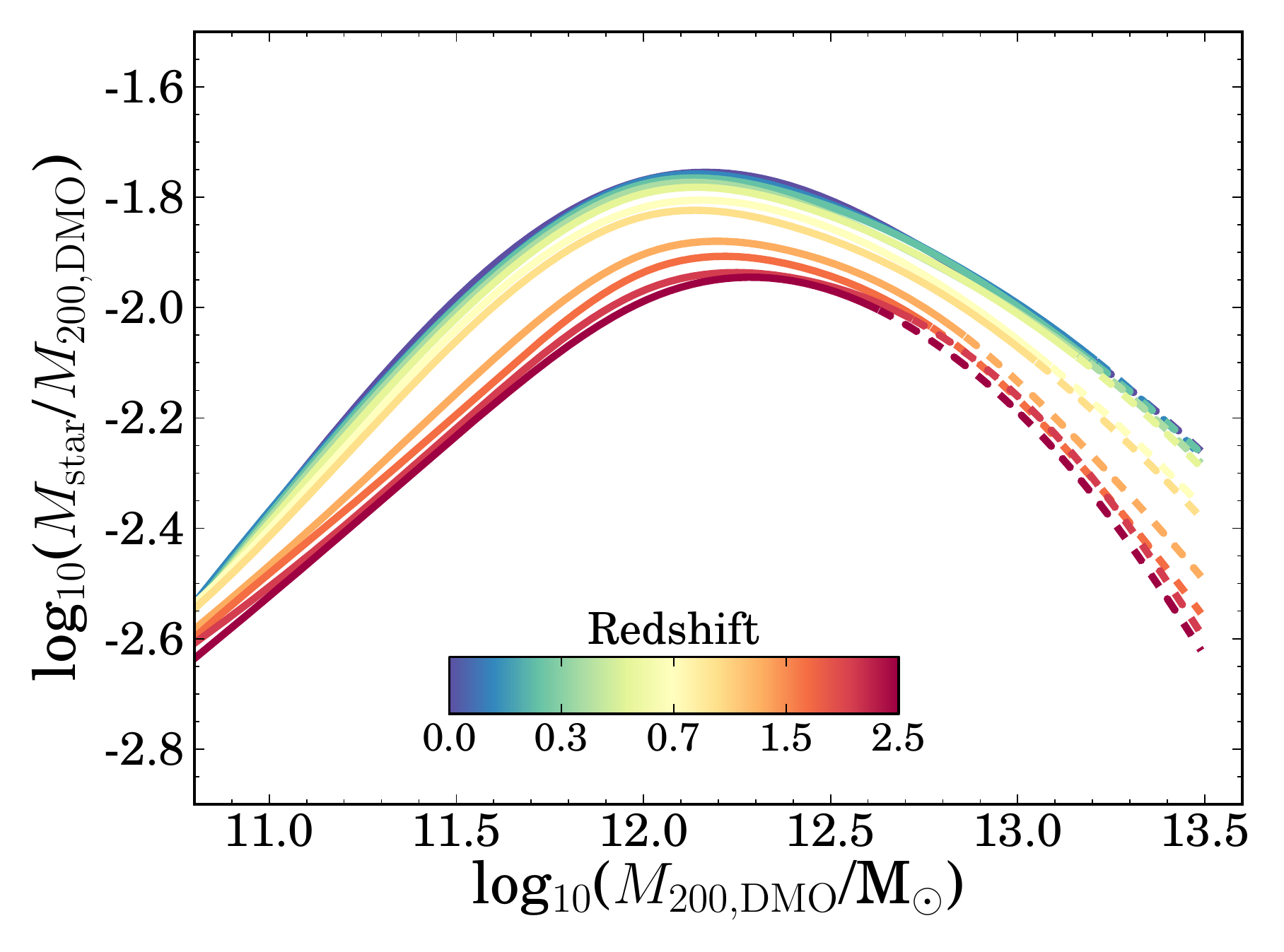}&
\includegraphics[width=8.6cm]{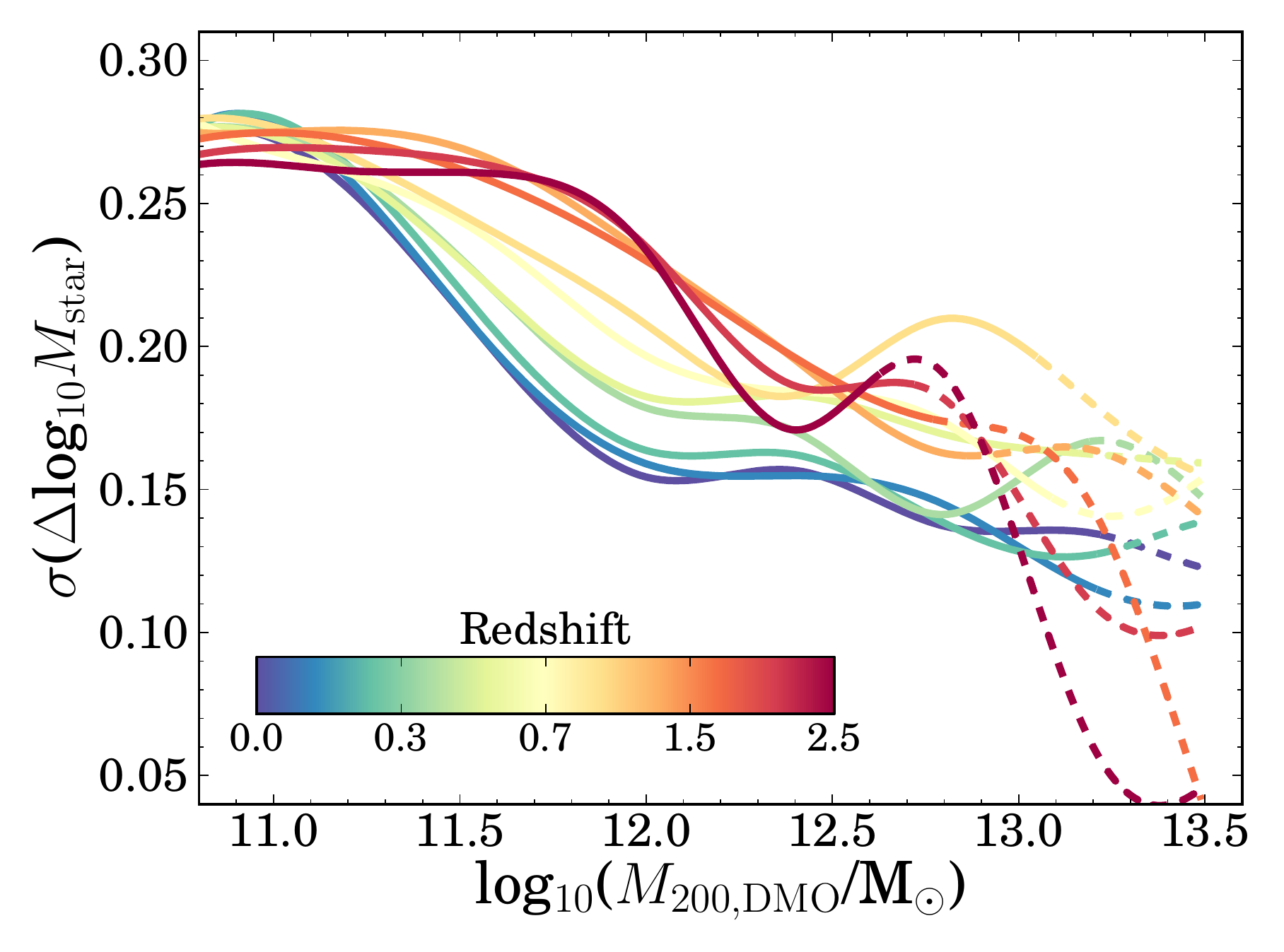}\\
\end{tabular}
\caption{\small{Evolution of the SMHM relation (left) and its scatter (right). Different from previous figures, we plot $M_{\rm star}$/$M_{200, \rm DMO}$ along the y-axis in order to increase the dynamic range. Dashed lines indicate where there are fewer than 100 galaxies per halo mass bin of 0.4 dex width. With increasing redshift the normalization of the SMHM drops and, except at the lowest halo masses, the scatter increases.}}
\label{fig:redshift_evo}
\end{figure*}

\section{Evolution}
In this section we investigate the evolution of the SMHM relation and the scatter in stellar mass as a function of halo mass. As we did for $z=0.1$, we fit the relation between stellar mass and $M_{200, \rm DMO}$ for central galaxies at different output redshifts from the EAGLE simulation using the non-parametric method.

We show $M_{\rm star}$/$M_{200, \rm DMO}$ versus $M_{200, \rm DMO}$ in the left panel of Fig. $\ref{fig:redshift_evo}$, as this better highlights the differences in comparison to showing stellar mass as a function of halo mass. There is almost no evolution between $z=0$ and $z=0.3$. At higher $z$, the evolution of the SMHM relation is, to first order, described by a decreasing normalisation: the ratio between stellar mass and halo mass decreases with increasing redshift, roughly independent of halo mass. At $M_{200, \rm DMO} \approx 10^{11}$ M$_{\odot}$, $M_{\rm star}$/$M_{200, \rm DMO}$ is roughly 0.15 dex lower at $z=2.5$ than at $z=0$. The evolution is largest at halo masses $\approx 10^{12}$ M$_{\odot}$, with M$_{\odot}$, $M_{\rm star}$/$M_{200, \rm DMO}$ decreasing by roughly 0.25 dex from $z=0$ to $z=2.5$. As a consequence, the peak in $M_{\rm star}$/$M_{200, \rm DMO}$ shifts to slightly higher masses with increasing redshift.

At fixed halo mass and $z = 0$ the normalisation of $M_{\rm star}$/$M_{200, \rm DMO}$ is about 0.2-0.3 dex lower in EAGLE than inferred from abundance matching by \cite{Behroozi2013} and \cite{Moster2013}. Similarly to \cite{Behroozi2013}, we find that the halo mass at which $M_{\rm star}$/$M_{200, \rm DMO}$ peaks increases only slightly ($\approx 0.1-0.2$ dex) between $z=0$ and $z=2.5$, while \cite{Moster2013} find a larger shift of $\approx0.6$ dex. Contrary to \cite{Behroozi2013} (who find a constant or even increasing peak $M_{\rm star}$/$M_{200, \rm DMO}$ with redshift), we find that the peak $M_{\rm star}$/$M_{200, \rm DMO}$ decreases by $\approx 0.2$ dex between $z = 0$ and $z = 2.5$, which is more similar to the trend found by \cite{Moster2013}.

In the right panel of Fig. $\ref{fig:redshift_evo}$, we show the evolution of the scatter in the SMHM relation between $z=0$ and $z=2.5$. While we find a relatively constant scatter for $M_{200, \rm DMO} \approx 10^{11}$ M$_{\odot}$, there is significantly more scatter for $M_{200, \rm DMO} \approx 10^{11.5-12.0}$ M$_{\odot}$ at higher redshifts. This could be the caused by a larger spread in halo formation times at these higher redshifts. The evolution of the scatter in the SMHM relation at higher halo masses is unconstrained due to uncertainties stemming from limited number statistics in the EAGLE volume.

\section{Discussion}
\subsection{Mass dependence of scatter}
As shown in Fig. $\ref{fig:scatter_mstar_haloproperties}$, the scatter in the difference between true stellar masses and stellar masses computed from fits to the SMHM relation, $\sigma(\Delta \rm log_{10} M_{\rm star})$, decreases with increasing halo mass, at least up to a halo mass of $M_{200, \rm DMO} \approx 10^{12}$ M$_{\odot}$. This is not a result of the limited volume of the EAGLE simulation, as shown in Appendix B. This is in contrast with the typical assumption that $\sigma({\rm log}_{10} M_{\rm star})$ is not a strong function of halo mass, and roughly equals 0.2 dex \citep[e.g.][]{Leauthaud2012,Moster2013,Uitert2016}, which is often used in halo abundance matching modelling \citep[e.g.][]{Behroozi2013}. However, as noted by \cite{Vakili2016}, abundance matching models that allow for assembly bias (see \S 7.2) indicate that the scatter can be significantly larger.

\begin{figure}
\centering
\includegraphics[width=8.6cm]{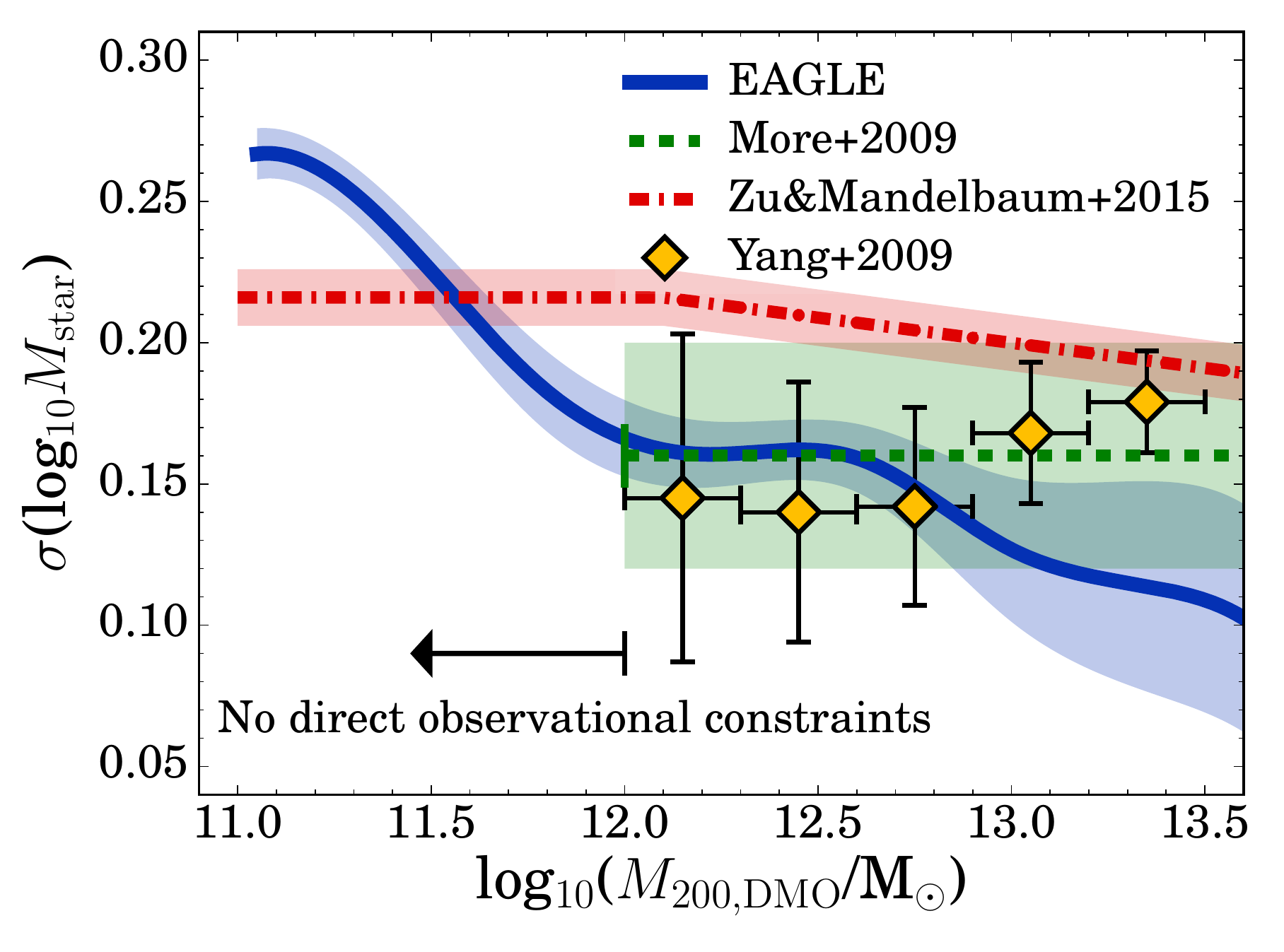}
\caption{\small{Scatter in the stellar mass at fixed halo mass as a function of halo mass. Yellow points show the binned results from the SDSS galaxy group catalog from \citealt{Yang2009b}. The green shaded region shows the observational constraints from satellite kinematics (\citealt{More2009b}). Both observational constraints are inferred for samples of galaxies with $M_{200, \rm DMO} > 10^{12}$M$_{\odot}$ and are consistent with the results from EAGLE for this mass range. The red dashed line and shaded region shows the mass dependent scatter inferred by \citet{ZuMandelbaum2015}. We note that $\sigma$($\Delta$log$_{10}$ $M_{\rm star}$) in EAGLE is intrinsic, and does not take errors in stellar masses into account, which do affect the observations. Therefore, the observational constraints should be considered as upper limits. }}
\label{fig:scatter_comparison}
\end{figure}

The most direct observational (yet model-dependent) constraints on the mass dependence of $\sigma(\rm log_{10} M_{\rm star})$ come from \cite{More2009b} and \cite{Yang2009b}, who both measure a halo mass independent scatter of $\approx 0.17$ dex. The observations from \cite{More2009b} are based on the kinematics of satellite galaxies, while \cite{Yang2009b} use a galaxy group catalog from the Sloan Digital Sky Survey (SDSS). However, for observational reasons, these constraints are mostly set at halo masses $M_{200, \rm DMO} > 10^{12}$ M$_{\odot}$ which are higher than the masses for which we find a significant trend. As illustrated in Fig. $\ref{fig:scatter_comparison}$, EAGLE is consistent with these observational constraints, contrarily to some semi-analytical models of galaxy formation, which produce much greater scatter \citep{Guo2016}. We note that the observational measurements of the scatter should be considered as upper limits due to errors in stellar mass measurements. 

By fitting to lensing and clustering measurements, \cite{ZuMandelbaum2015} simultaneously constrain the SMHM relation and its scatter at $z\sim0.1$. In agreement with our results, they find that $\sigma(\rm log_{10} M_{\rm star})$ decreases with increasing halo mass: from 0.22$^{+0.02}_{-0.01}$ dex at $M_{200} \lesssim 10^{12}$ M$_{\odot}$ to  0.18$^{+0.01}_{-0.01}$ dex at $M_{200} \approx 10^{14}$ M$_{\odot}$ (see also Fig. $\ref{fig:scatter_comparison}$ and the results from a semi-empirical approach by \citealt{Rodriguez-Puebla2015}). This scatter is similar to that we find in EAGLE at $M_{200} \lesssim 10^{12}$ M$_{\odot}$, but it decreases more slowly with halo mass than in EAGLE, which may be due to the constraints set by their assumption that the mass dependence of the scatter follows a simple linear relation. \cite{Tinker2016a} report a $0.18^{+0.01}_{-0.02}$ dex scatter at $M_{200} \gtrsim 10^{12.7}$ M$_{\odot}$, slightly higher than the scatter in the SMHM in EAGLE. Combined with the result from \cite{Kravtsov2014} (who find that the scatter is $0.17\pm0.02$ dex for $M_{200}>10^{14}$ M$_{\odot}$), the observations indicate that the scatter in the SMHM relation is insensitive of halo mass for haloes more massive than $M_{200} \gtrsim 10^{12.7}$ M$_{\odot}$.

In a recent combined analysis of $N$-body simulations and the fitted SMHM relation from \cite{Behroozi2013_SMHM}, \cite{Gu2016} study the origin of scatter in the SMHM relation. Although their analysis is limited to haloes with masses $M_{200, \rm DMO}>10^{12}$M$_{\odot}$, they argue that the constant scatter in the SMHM relation as a function of halo mass is due to an interplay of scatter due to ex-situ growth (i.e. accretion) and in-situ growth (i.e. star formation), and that the observed independence of the scatter in the SMHM relation on halo mass is a coincidence. In the analysis of \cite{Gu2016}, hierarchical assembly leads to a scatter of $\approx 0.16$ dex at high halo masses, which is roughly independent of the details of galaxy formation. Although the scatter due to ex-situ growth increases towards low halo masses, the relative importance of in-situ growth dominates below $M_{200, \rm DMO}<10^{12}$M$_{\odot}$. Therefore, the scatter in the SMHM relation at lower masses is set mostly by the scatter in the in-situ growth at fixed halo mass, which is more strongly related to the details of galaxy formation. \cite{Gu2016} assume the scatter in in-situ mass growth to be 0.2 dex for all halo masses. However, if this scatter were higher, or increases with decreasing halo mass, the resulting scatter in the SMHM relation will increase with decreasing halo mass. This would be consistent with our findings based on the EAGLE simulation, particularly since we find that at low halo masses the scatter in the SMHM relation depends strongly on halo formation time (which is likely related to in-situ star formation or binding energy).

More observational constraints on the scatter in the SMHM relation at low halo masses would be valuable. However, it is observationally challenging to measure halo masses for lower-mass central galaxies using methods such as satellite kinematics or galaxy-galaxy lensing \citep[e.g.][]{Mandelbaum2006,ZuMandelbaum2015}. At lower stellar masses, halo masses may be estimated from measured rotational velocities (e.g. \citealt{Blanton2008,Reddick2013} and see for example the compilation by \citealt{Leauthaud2012}). However, adiabatic contraction due to galaxy formation increases halo concentrations and thus also rotational velocities at fixed halo mass \citep[e.g.][]{Desmond2015}. Thus, the measured scatter in stellar mass at fixed halo mass would be biased low if halo masses were based on rotational velocities.

\subsection{Relation to assembly bias \& abundance matching}
A commonly used technique to connect stellar masses to halo masses is abundance matching, which, in its simplest form, assumes that stellar mass increases monotonically with halo mass (or another halo property such as the maximum rotational velocity). This is related to the assumption that the halo clustering strength is fully determined by the halo mass. It may be plausible that a second halo property that is related to halo clustering is also a second parameter in determining galaxy stellar masses. 

The existence of a second halo property that is related to the clustering strength of haloes is called assembly bias \citep[e.g.][]{Gao2005,Dalal2008, LacernaPadilla2011}. These studies have shown that the clustering of dark matter haloes depends not only on halo mass, but also on their formation time. \cite{Chaves2015} showed that assembly bias significantly alters the clustering of galaxies in the EAGLE simulation, and that using $V_{\rm relax}$ as a halo property suffers significantly less from this effect. There is no consensus in observations of large sets of galaxies (such as 2dFGRS or SDSS) on the existence of assembly bias. While some authors claim a signal from assembly bias \citep{Yang2006,Wang2008}, others \citep{BlantonBerlind2007,Tinker2008} argue little to no dependence of (for example) galaxy colour on large scale clustering \citep[c.f.][]{Hearin2016}. More recently, \cite{Zentner2016} and \cite{Vakili2016} argue that significant assembly bias cannot be excluded from galaxy clustering data from SDSS.

The existence of scatter in the SMHM relation means that halo mass alone cannot predict stellar mass with an accuracy better than $\approx0.2$ dex, although we note again that this scatter decreases with increasing halo mass, see e.g. Fig. $\ref{fig:mstarreconstruct}$. The scatter in stellar masses can be slightly reduced by using information about the concentration or formation time of the haloes. This is similar to the approach by \cite{Lehmann2015} and \cite{Hearin2016}, who extend the abundance matching method to halo properties that also depend on halo concentration. Thus, halo properties such as $V_{\rm max}$ or binding energy, which are related to both halo mass and concentration, are the most fundamental halo properties in determining stellar masses (see also \citealt{Reddick2013}). This resembles the conclusion from \cite{BoothSchaye2010}, who argue that the halo binding energy determines black hole mass and indicates a co-evolution of galaxies and their massive black holes.
 
From an analysis of local galaxies from the SDSS, \cite{Lim2015} argued that there is a relation between the ratio of a central galaxy's stellar mass to its halo mass (from the \citealt{Yang2009b} group catalog) and the galaxy formation time. They find that galaxies with a high ratio of central stellar mass to halo mass are typically redder, older and more bulge-like: properties that are all associated with older stellar populations. In EAGLE, as illustrated in Fig. $\ref{fig:SMHM_colorz05}$, such a relation also exists between DMO halo formation time and the ratio of stellar to halo mass. In this case, it certainly indicates a causal relationship, since the halo formation time is measured in the independent DMO version of EAGLE.

\subsection{The origin of the remaining scatter}
Intriguingly, we find that there remains significant scatter in the SMHM relation after accounting for the effect of concentration, and that there is also scatter in the $M_{\rm star}- V_{\rm max}$ relation, which is not strongly related to any of the dimensionless DMO halo properties (see \S 4.2) studied. We consider possible explanations for the remaining scatter: 
\begin{enumerate}
\item The scatter reflects noise due to the finite numerical resolution. We think this is unlikely since particularly the highest-mass galaxies are resolved using $>10,000$ particles. In appendix B we compare with a higher-resolution 25 Mpc EAGLE simulation. Although this comparison is only possible for a volume that is too small to sample halo masses $\gtrsim10^{12.5}$ M$_{\odot}$, we find no evidence for significant resolution effects.
\item The scatter is caused by a combination of weak correlations with halo properties that are uncorrelated with the halo properties we included. In particular, baryonic processes might be very non-linear and chaotic, such that only small differences in halo properties result in a substantial differences in the final stellar mass. Examples may include rare but explosive feedback or the interplay between dissipation, collapse and feedback.
\end{enumerate}

\section{Conclusions}
We have used the EAGLE cosmological hydrodynamical simulation to study what drives the efficiency of galaxy formation in halos hosting central galaxies. In particular, we studied which dimensional dark matter halo property $X$ is most closely related to stellar mass, and whether other dimensionless halo properties are responsible for driving the scatter in the stellar mass at fixed $X$ for halo masses from $10^{11-13.5}$ M$_{\odot}$ (corresponding to $M_{\rm star} \approx 10^{8-11.5}$ M$_{\odot}$). The investigated dimensional dark matter halo properties include different definitions of halo mass, binding energy and rotational velocity, see Table $\ref{tab:properties}$, while the investigated dimensionless halo properties are concentration, formation time, spin, sphericity, triaxiality, environment and substructure.

Since differences between haloes are ultimately determined by differences in the initial conditions, dominated by the initial density perturbations of (primarily) dark matter, we used halo properties from a matched DMO simulation. This is necessary since properties in the baryonic EAGLE simulation are affected by baryonic processes, making cause and effect impossible to separate. For example, if the baryonic simulations predicts a correlation between stellar mass and dark matter halo concentration, then it is not clear whether a higher concentration causes a higher stellar mass or vice versa. \\
The main conclusions of this work are:

\begin{enumerate}
\item The maximum circular velocity, $V_{\rm max, DMO}$, is the DMO halo property that is most closely related to stellar mass. The binding energy measured at $R_{2500, \rm DMO}$ is almost as strongly correlated with stellar mass. These halo properties are more fundamental than $M_{200, \rm DMO}$, for which there is more scatter in stellar mass, see Figs. $\ref{fig:scatter_mstar_haloproperties}$ and $\ref{fig:mhalo_variations}$. We have provided formulae to compute stellar masses based on dark matter halo properties (\S 5).

\item The scatter in stellar mass at fixed halo mass decreases with increasing halo mass, from $\approx0.25$ dex at $M_{200, \rm DMO} = 10^{11}$ M$_{\odot}$ to $\approx0.12$ dex at $M_{200, \rm DMO} = 10^{13}$ M$_{\odot}$. This is in contrast with the common assumption that the amount of scatter is independent of halo mass. 

\item For halo masses $M_{200, \rm DMO} > 10^{12}$ M$_{\odot}$ the scatter in stellar mass decreases much less rapidly with halo mass than for $M_{200, \rm DMO} < 10^{12}$ M$_{\odot}$. This may be due to the transition from feedback dominated by star formation to feedback dominated by AGN, or due to the increased importance of mergers.

\item The measured scatter at $M_{200, \rm DMO} > 10^{12}$ M$_{\odot}$ is consistent with the most direct inferences from observations (Fig. $\ref{fig:scatter_comparison}$). Future direct observations probing lower halo masses can test whether there is indeed more scatter in stellar mass at low halo mass.

\item The halo concentration, which is a good proxy for formation time, is responsible for part of the scatter in the stellar mass - halo mass relation up to a halo mass of $M_{200, \rm DMO}  \sim 10^{12.5}$ M$_{\odot}$, see Figs. $\ref{fig:smhm_res_c200}$ and $\ref{fig:range_spearman}$. Haloes with a higher concentration formed earlier and have been more efficient at forming stars (Fig. $\ref{fig:SMHM_colorz05}$), probably because they are in a more advanced stage of their evolution and/or because it is harder for feedback to drive winds out of haloes with a higher concentration. 

\item By correcting for concentration or formation time using a functional form, the scatter in the SMHM relation can be reduced by up to 0.04 dex (depending on the halo mass range), see Fig. $\ref{fig:range_spearman}$. However, the remaining scatter in stellar mass is as large as the scatter in the $M_{\rm star}$ - $V_{\rm max}$ relation, see Fig. $\ref{fig:mstarreconstruct}$.

\item Empirical models, which assign stellar masses to haloes in DMO simulations are more reliable if halo properties based on both halo mass and concentration are used, such as $V_{\rm max}$. 

\item We find no DMO halo property that can account for the scatter in the SMHM relation after correcting for the effect of concentration (or formation time), or for the scatter the $M_{\rm star}- V_{\rm max}$ relation. This means that, except for properties related to the halo mass and concentration, other halo properties (such as spin, sphericity, triaxiality, environment and substructure) are statistically unimportant for determining the stellar mass of a galaxy. It is therefore likely that more complex (combinations of) halo properties and assembly histories are responsible for the remaining scatter in stellar masses by driving chaotic non-linear baryonic effects.

\item There is little evolution in the SMHM relation between $z=0$ and $z=0.3$. At higher redshift the evolution of the SMHM relation is approximately described by a decreasing normalisation, relatively independent of halo mass. The evolution is largest at halo masses $\approx 10^{12}$ M$_{\odot}$. As a consequence, the peak in $M_{\rm star}$/$M_{200, \rm DMO}$ shifts to slightly higher masses with increasing redshift. While the scatter in the SMHM is relatively constant between $z=0$ and $z=2.5$ for haloes with $M_{200, \rm DMO} \approx 10^{11}$ M$_{\odot}$, we find that there is significantly more scatter for $M_{200, \rm DMO} \approx 10^{11.5-12.0}$ M$_{\odot}$ at $z\gtrsim1$ than at $z\approx0$ (see Fig. $\ref{fig:redshift_evo}$). This is likely caused by the larger spread in halo formation times at these higher redshifts.
\end{enumerate}

The efficiency of galaxy formation, defined as the scatter in the stellar mass - halo mass relation for central galaxies, is determined by the cosmological initial conditions. Haloes which reside in more over-dense regions collapse earlier, leading to a higher concentration and an earlier formation of stars and less efficient feedback due to deeper potential wells. Measures of the potential well depth, such as $V_{\rm max}$, $M_{2500, \rm DMO}$ or a combination of these properties, correlate more strongly with stellar mass than $M_{200, \rm DMO}$ alone and are thus more fundamental properties governing the evolution of galaxies. However, this is only valid up to halo masses of $\sim10^{12.5}$ M$_{\odot}$. Beyond this mass, additional physical processes play a role, as a more concentrated halo does not necessarily lead to a higher stellar mass. 

\section*{Acknowledgments}
We thank the anonymous referee for their comments. JM acknowledges the support of a Huygens PhD fellowship from Leiden University. JM thanks David Sobral for useful discussions and the help with fitting routines and Jonas Chavez-Montero and Ying Zu for providing data. We thank PRACE for the access to the Curie facility in France. We have used the DiRAC system which is a part of National E-Infrastructure at Durham University, operated by the Institute for Computational Cosmology on behalf of the STFC
DiRAC HPC Facility (www.dirac.ac.uk); the equipment was funded by BIS National E-infrastructure capital grant ST/K00042X/1, STFC capital grant ST/H008519/1, STFC DiRAC Operations grant ST/K003267/1 and Durham University.
The study was sponsored by the Dutch National Computing Facilities Foundation (NCF) for the use of supercomputer facilities, with financial support from the Netherlands Organisation for Scientific Research (NWO), through VICI grant 639.043.409, and the European Research Council under the European Union's Seventh Framework Programme (FP7/2007- 2013) / ERC Grant agreement 278594-GasAroundGalaxies, and from the Belgian Science Policy Office ([AP P7/08 CHARM]). RAC is a Royal Society University Research Fellow.
We have benefited greatly from the public available programming language {\sc Python}, including the {\sc numpy, matplotlib, pyfits, scipy, h5py} and {\sc rpy2} packages, and the {\sc Topcat} analysis program \citep{Topcat}.




\bibliographystyle{mnras}

\bibliography{bibliography_pceagle.bib}


\appendix
\section{Varying the definition of stellar mass}

\begin{figure}
\includegraphics[width=8.5cm]{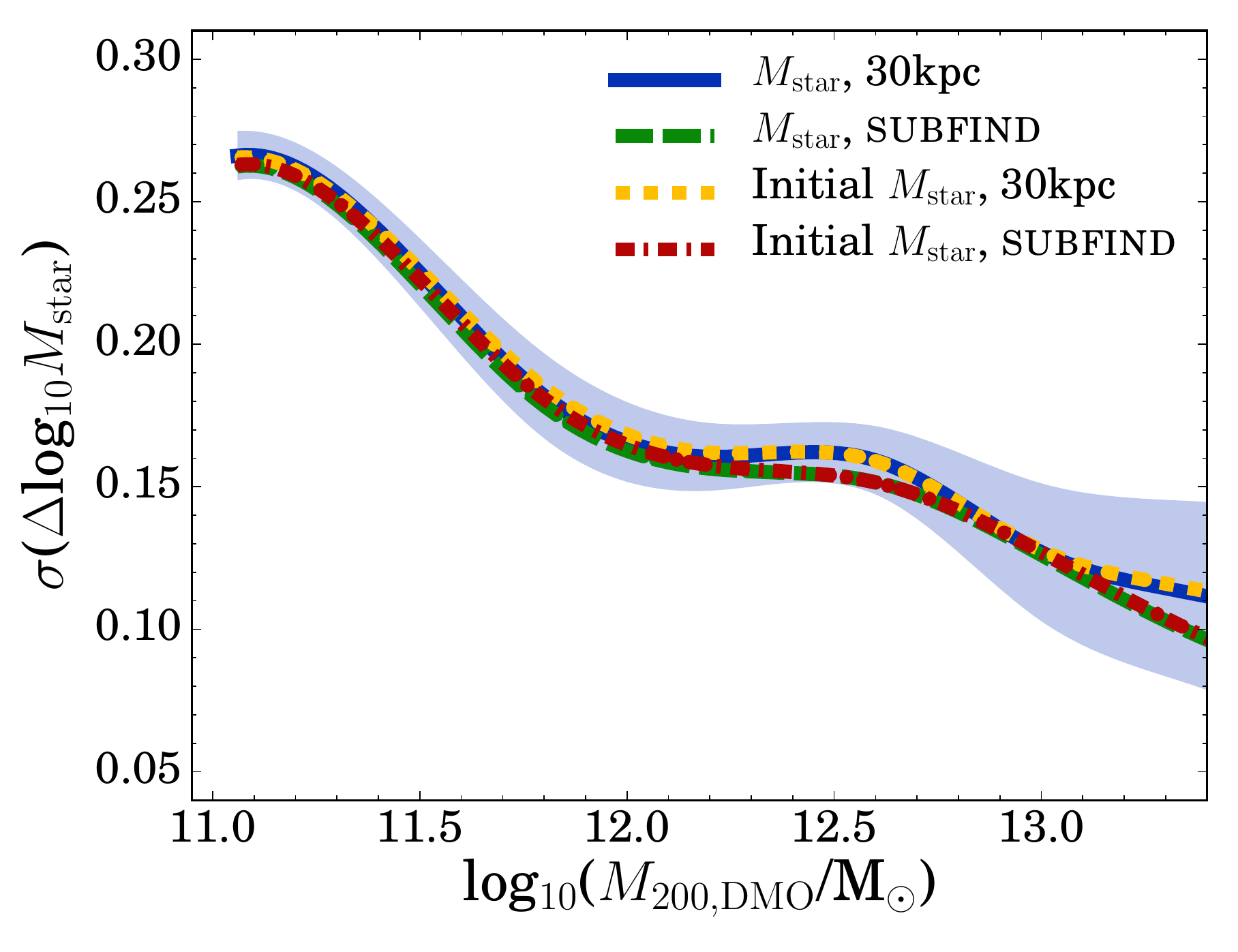}\\
\caption{The effect of scatter in stellar masses for varying definitions of stellar mass. The total stellar mass in a subhalo is slightly more closely related to halo mass than the aperture stellar mass is. There is negligible difference between the scatter in initial (i.e. corrected for stellar mass loss due to stellar evolution) and current with initial stellar mass.}
\label{fig:scatter_mstar_variations}
\end{figure}

In this appendix, we vary the definition of stellar mass. In the main text, we used the total mass of star particles within 30 proper kpc of the minimum of the gravitational potential of a subhalo at $z=0.1$. However, high-mass haloes contain substantial stellar mass at larger radii. We therefore use the stellar mass of all particles in the subhalo as identified by {\sc subfind}. As illustrated in Fig. $\ref{fig:scatter_mstar_variations}$, the spread in stellar mass as a function of halo mass becomes slightly lower for the highest halo masses. 

Another variation that we investigate is the initial stellar mass (either within 30 kpc or of all star particles in the {\sc subfind} subhalo). This is the mass that a stellar particle had at the time it was formed, and using this mass removes the effects of stellar mass loss. Fig. $\ref{fig:scatter_mstar_variations}$ shows no significant differences when we use this definition. Although the typical stellar mass loss is $40-50$ \%, the difference in stellar mass loss between the youngest and oldest galaxies is small, only $\approx20$ \%, as the majority of stellar mass loss occurs on timescales $<10^9$ yr due to the limited lifetimes of massive stars \citep[e.g.][]{Segers2015}. When using the initial stellar mass, the correlations between the scatter in the SMHM relation and concentration/formation time become slightly stronger ($\sim + 0.02$ Spearman rank). This is easy to understand: the most concentrated haloes form the earliest, such that the effect of stellar mass loss will be highest, and this will therefore slightly weaken the trend that an earlier formation time leads to a higher redshift zero stellar mass (at fixed present-day halo mass).

\begin{figure}
\includegraphics[width=8.5cm]{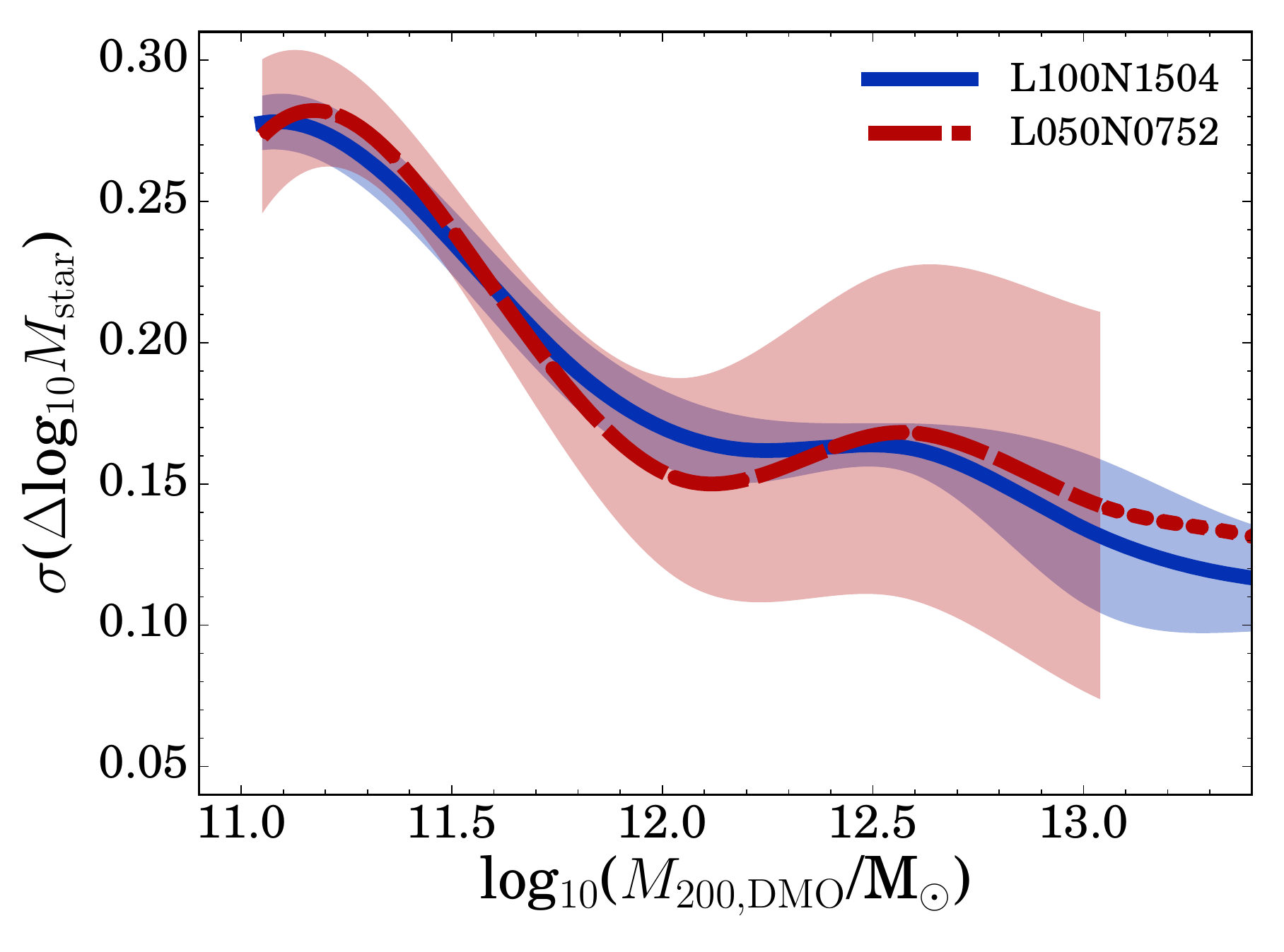}\\
\includegraphics[width=8.5cm]{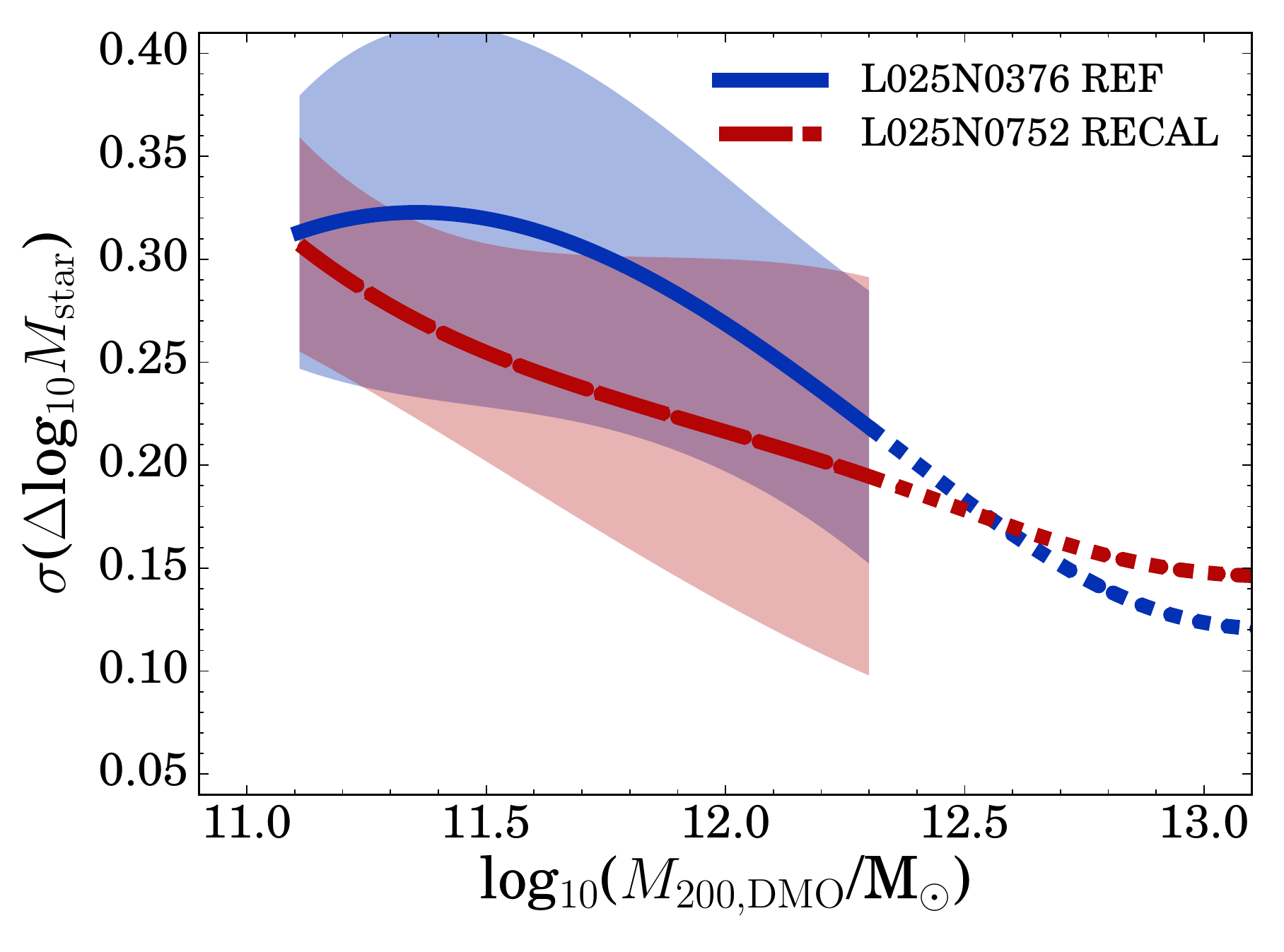}
\caption{\small{Box size and resolution test. Scatter in the difference between true stellar masses (in the baryonic simulation) and stellar masses computed from the non-parametric fits to the relation between stellar mass and DMO halo mass, as a function of DMO halo mass. The errors are estimated by jackknife resampling the data in sub-volumes with 1/8th times the total volume of the box. For halo mass ranges where there are less than 25 haloes per bin, we do not show the errors and only indicate the relation with a dashed line.}}
\label{fig:box_res_test}
\end{figure} 

\section{Dependence on box size and resolution}
In this appendix, we test whether the scatter in the stellar mass - halo mass relation as a function of halo mass depends on the simulated volume or resolution. For the box size test, we compare the results from the (100 Mpc)$^3$ box with those from a (50 Mpc)$^3$ box with the same resolution. As shown in the top panel of Fig. $\ref{fig:box_res_test}$, the spread in stellar mass as a function of halo mass is very similar between the two boxes. This means that our conclusions are not affected by the finite volume of our simulation. Small number statistics (in terms of number of galaxies and in terms of independent environments in the simulation volume) does increase the uncertainty in $\sigma$($\Delta$log$_{10}$ $M_{\rm star}$) with halo mass. The top panel of Fig. $\ref{fig:box_res_test}$ shows that this effect is much stronger in the smaller box, such that the increase in uncertainty with mass may be due to the finite box size.

In the bottom panel of Fig. $\ref{fig:box_res_test}$, we compare the spread in stellar mass as a function of halo mass in two simulations with a box size of (25 Mpc)$^3$. Note that we had to increase the bin-widths from 0.4 to 0.6 dex in order not to be dominated by errors. While one simulation has the {\it fiducial} resolution (L025N0376), the other uses a spatial (mass) resolution better by a factor two (eight). We note that the Recal model parameters differ slightly from those of the Reference model, see \cite{Schaye2014}. It is hard to reliably investigate the effect of resolution on the statistical scatter in stellar mass, because the errors are very large due to the small numbers of galaxies per halo mass bin. While the differences are within the error bars, there might be less scatter in stellar mass at fixed halo mass in the simulation with higher resolution, particularly for $M_{200, \rm DMO} < 10^{12.5}$ M$_{\odot}$.


\bsp	
\label{lastpage}
\end{document}